\documentclass[12pt,letterpaper]{article}  

\usepackage[includeheadfoot,
            marginratio={1:1,2:3}, 
            width=412pt, 
            height=688pt,]{geometry}

\usepackage{amsmath}
\usepackage{amsfonts}
\usepackage{amssymb}
\usepackage{stmaryrd}
\usepackage{dsfont}
\usepackage{graphicx}
\usepackage{cite}

\newcommand{\eq}[1]{\begin{equation}
                     \begin{split} #1 \end{split}
                     \end{equation}}

\newcommand{\ul}{\underline}
\newcommand{\ov}{\overline}
\newcommand{\lab}{\mathsf }
\newcommand{\op}{\hspace{1pt}}
\newcommand{\slnabla}{\slash\hspace{-7pt}\nabla}

\allowdisplaybreaks[2]
\numberwithin{equation}{section}
\renewcommand*{\thefootnote}{\fnsymbol{footnote}}

\begin{document}

\vspace*{-1.5cm}
\begin{flushright}
  {\small
  MPP-2015-177\\
  LMU-ASC 49/15\\
  }
\end{flushright}

\vspace{4em}

\begin{center}
{\LARGE
Relating Double Field Theory to the Scalar Potential of
 N=2 Gauged Supergravity \\[0.3cm]
}
\end{center}

\vspace{1em}

\begin{center}
Ralph Blumenhagen$^{1}$, Anamaria Font$^{1,2}$\footnote{On leave from Departamento de F\'{\i}sica, Facultad de Ciencias, Universidad Central de Venezuela}, Erik Plauschinn$^{2}$
\end{center}

\vspace{1em}

\begin{center} 
\emph{
$^{1}$ Max-Planck-Institut f\"ur Physik (Werner-Heisenberg-Institut), \\ 
F\"ohringer Ring 6,  80805 M\"unchen, Germany 
} \\[1em] 

\emph{
$^{2}$ Arnold Sommerfeld Center for Theoretical Physics,\\ 
LMU, Theresienstr.~37, 80333 M\"unchen, Germany
}
\end{center} 

\vspace{3em}

%%%%%%%%%%%%%%%%%%%%%%%%%%%%%%%%%%%%%%%%%%%%%%%
%%%%%%%%%%%%%%%%%%%%%%%%%%%%%%%%%%%%%%%%%%%%%%%
%%%%%%%%%%%%%%%%%%%%%%%%%%%%%%%%%%%%%%%%%%%%%%%
%%%%%%%%%%%%%%%%%%%%%%%%%%%%%%%%%%%%%%%%%%%%%%%
%%%%%%%%%%%%%%%%%%%%%%%%%%%%%%%%%%%%%%%%%%%%%%%
%%%%%%%%%%%%%%%%%%%%%%%%%%%%%%%%%%%%%%%%%%%%%%%
%%%%%%%%%%%%%%%%%%%%%%%%%%%%%%%%%%%%%%%%%%%%%%%
%%%%%%%%%%%%%%%%%%%%%%%%%%%%%%%%%%%%%%%%%%%%%%%

\renewcommand*{\thefootnote}{\arabic{footnote}}
\setcounter{footnote}{0}

\begin{abstract}
The double field theory action in the flux formulation is dimensionally reduced on a Calabi-Yau
three-fold equipped with non-vanishing type IIB geometric and non-geometric
fluxes. First, we rewrite the metric-dependent reduced DFT action in terms
of quantities that can be evaluated without explicitly knowing the metric
on the Calabi-Yau manifold. Second, using properties of special geometry we 
obtain the scalar potential of N=2 gauged supergravity. After
an orientifold projection, this potential is consistent with the scalar potential
arising from the flux-induced superpotential, plus
an additional D-term contribution. 
\end{abstract}

\clearpage

\tableofcontents

%%%%%%%%%%%%%%%%%%%%%%%%%%%%%%%%%%%%%%%%%%%%%%%
%%%%%%%%%%%%%%%%%%%%%%%%%%%%%%%%%%%%%%%%%%%%%%%
%%%%%%%%%%%%%%%%%%%%%%%%%%%%%%%%%%%%%%%%%%%%%%%
%%%%%%%%%%%%%%%%%%%%%%%%%%%%%%%%%%%%%%%%%%%%%%%
%%%%%%%%%%%%%%%%%%%%%%%%%%%%%%%%%%%%%%%%%%%%%%%
%%%%%%%%%%%%%%%%%%%%%%%%%%%%%%%%%%%%%%%%%%%%%%%
%%%%%%%%%%%%%%%%%%%%%%%%%%%%%%%%%%%%%%%%%%%%%%%
%%%%%%%%%%%%%%%%%%%%%%%%%%%%%%%%%%%%%%%%%%%%%%%

\section{Introduction}
\label{sec:intro}

One of the main issues in relating string theory to observable
physics is the problem of moduli stabilization. 
For instance supersymmetric compactifications of string theory
on Calabi-Yau (CY) manifolds come with a plethora of massless scalars,
so-called moduli. At string tree-level these moduli can be stabilized
by turning on  fluxes on the Calabi-Yau manifold 
(for reviews see e.g.\cite{Grana:2005jc,Douglas:2006es,Denef:2007pq}).
This procedure is mostly discussed in an effective four-dimensional
framework, i.e. one starts with the initial CY geometry and
considers the fluxes as off-shell deformations of the theory.
In the effective description this leads to the generation of a scalar
potential that depends on the moduli fields. The hope is that
new classical field theory vacua of this scalar potential reflect 
new solutions to the true ten-dimensional equations of motion. 
Finding these solutions concretely is a highly non-trivial step,
as it involves going away from the initial CY geometry.

The prime example of the application of flux-induced potentials  are type IIB models with non-trivial
Neveu-Schwarz--Neveu-Schwarz (NS-NS) and Ramond-Ramond (R-R)
three-form fluxes \cite{Taylor:1999ii,Giddings:2001yu}. 
In this case a no-scale
potential involving only the complex structure moduli and 
the axio-dilaton is obtained. The K\"ahler moduli remain massless, but
can be lifted by subleading perturbative and non-perturbative effects. 
This is the idea behind the KKLT \cite{Kachru:2003aw} and LARGE volume scenario \cite{Balasubramanian:2005zx}.

It is known that in order to stabilize also the K\"ahler moduli at tree-level,
one needs to  consider additional non-geometric fluxes
\cite{Shelton:2005cf,Wecht:2007wu}.  For the toroidal case this has been
investigated in
\cite{Aldazabal:2006up,Villadoro:2006ia,Shelton:2006fd,deCarlos:2009fq, Aldazabal:2011yz,Dibitetto:2011gm} 
and for generic Calabi-Yau manifolds
in \cite{Grana:2005ny,Benmachiche:2006df,Grana:2006hr,Micu:2007rd,Palti:2007pm,Cassani:2007pq, Robbins:2007yv}, 
among others.
In particular, it has been shown that  the generalized
flux-induced scalar potential 
can be related to the scalar potential of  N=2 gauged supergravity
\cite{D'Auria:2007ay}. Lately, this kind of flux vacua have
been  investigated from a string phenomenological point of view, with
special emphasis on realizing F-term axion monodromy inflation \cite{Blumenhagen:2015kja}.

From the higher dimensional point of view, it has been argued that 
non-geometric aspects of string theory can be captured by
double field theory (DFT)\cite{Siegel:1993xq,Siegel:1993th,hull:2009mi,hohm:2010pp,hohm:2010jy}; for reviews see
\cite{Aldazabal:2013sca,Berman:2013eva,Hohm:2013bwa}. 
DFT provides a self-consistent
framework  that features new symmetries such as generalized diffeomorphisms
and a manifest global $O(D,D)$ symmetry that close upon
invoking the so-called strong constraint. In particular, though
derived from string field theory on a torus, DFT is claimed to
be background independent. See also \cite{Blumenhagen:2014gva,
Blumenhagen:2015zma} for the derivation
of a  DFT action resulting from string field theory
on  WZW models.

There exist two formulations of the DFT action, 
which differ by terms that  are either total derivatives or
vanish due to the strong constraint. For our purposes it is
convenient to use the  so-called flux formulation of the DFT action
in the form presented in  \cite{Geissbuhler:2013uka}.
This  is motivated by the scalar potential in gauged supergravity
which, as
shown in \cite{Hohm:2010xe}, is also related to the early work of 
W.~Siegel \cite{Siegel:1993xq,Siegel:1993th}.

It has been shown that compactifying or Scherk-Schwarz reducing DFT
on a toroidal background equipped with
constant geometric and non-geometric fluxes gives the scalar potential
of half-maximal gauged supergravity in four dimensions\cite{Aldazabal:2011nj,Geissbuhler:2011mx,Grana:2012rr}. 
The relation of DFT to the scalar potential of N=2 gauged
supergravity has however not explicitly been clarified.
Clearly, the expectation is that dimensionally reducing DFT on a 
genuine Calabi-Yau manifold carrying constant geometric and
non-geometric fluxes should give the scalar potential of N=2 gauged 
supergravity.
It is the purpose of this paper to fill this gap and explicitly show
how the dimensional reduction of DFT can be performed in order 
to match N=2 gauged supergravity results. This can be
considered as the generalization of the computation first performed in
\cite{Taylor:1999ii},
where the dimensional reduction of the kinetic terms of the NS-NS and
R-R type IIB three-form on a (non-toroidal) CY three-fold gives the no-scale scalar potential described
in supergravity language by the tree-level K\"ahler potential for the complex
structure/axio-dilaton moduli and the Gukov-Vafa-Witten (GVW)
superpotential \cite{Gukov:1999ya}.

The main technical
problem is that the action of DFT  contains the metric on the CY
three-fold, which is not explicitly known. Therefore, one first has to
appropriately rewrite the DFT action so that only quantities appear
that can be treated without knowing the metric explicitly. 
For instance, for the simple H-flux case we can write
\eq{
\label{heasy}
  \star\op{\cal L}=-\frac{e^{-2\phi}}{ 2} \op d^{10}x \op\sqrt{-G} \, H_{ijk}\, H_{i'j'k'}\, g^{ii'}
    g^{jj'} g^{kk'}=-{e^{-2\phi}\over 2}\op  H\wedge \star\op H\,,
}
but DFT contains many more terms that are not of this simple type.
To perform the dimensional reduction, it is most appropriate to
start with  the DFT action in the 
flux formulation.
This action essentially contains the kinetic terms of the various
geometric and non-geometric fluxes. We will treat the background fluxes
as constant parameters that are only subject to their Bianchi
identities, which are quadratic constraints for the constant fluxes.
The indices for these fluxes are contracted using the constant $O(D,D)$
metric or the generalized
metric. The latter  contains the background CY metric, hence depending on the complex
structure and complexified K\"ahler moduli. 
The CY metric is assumed to only depend on the usual 
coordinates.\footnote{The same
procedure to carry out the dimensional reduction/oxidation of DFT compactified on a torus (orbifold)  background was employed in
\cite{Blumenhagen:2013hva,Shukla:2015bca}.}

This paper is organized as follows: In section~\ref{sec_review} we provide  a brief
review of the main aspects of DFT that are relevant for this paper.
As mentioned,  we focus on the DFT action in the flux formulation.
In section~\ref{sec_dft_on_cy}, following a step by step procedure, we rewrite the 
action compactified on a CY in terms of quantities that only contain operations like
wedge products, the Hodge-star map and actions of fluxes
on $p$-forms.  The main result is the generalization of \eqref{heasy}. We find that
all NS-NS terms appearing in the DFT action can be rewritten as
\eq{
\label{onlyyou}
  \star \mathcal L_{\rm NS\op NS} = -e^{-2\phi}\bigg[\hspace{18pt}
  &
  {1\over 2}\, \chi \wedge \star\op \ov\chi \;+\;{1\over 2}\op \Psi\wedge \star\op \ov\Psi \\
  -& {1\over 4}\op \big(\Omega\wedge \chi\big)\wedge\star \op \big(\ov\Omega\wedge \ov \chi\big)
  -{1\over 4}\op \big(\Omega\wedge \ov\chi\big)\wedge\star\op \big(\ov\Omega\wedge  \chi\big)
  \hspace{8pt}\biggr]\,,
}
where $\chi={\mathfrak D} \op  e^{iJ}$ and $\Psi=\mathfrak D\op
\Omega$. Here ${\mathfrak D}$ denotes a twisted differential 
that involves all  geometric and non-geometric fluxes,
and $J$ and $\Omega$ are the K\"ahler  and holomorphic three-form 
of the CY.  
In section~\ref{sec_reduce}, this action is evaluated after
introducing the fluxes on the internal CY manifold as in \cite{Grana:2005ny,Grana:2006hr}. We show
that the resulting scalar potential takes the same form 
as the one in \cite{D'Auria:2007ay}, which was shown there to be equivalent to
the N=2 gauged supergravity result.  In section~\ref{sec_orient}, we perform an orientifold
projection of type IIB with O7-/O3-planes and show that the scalar
potential derived from \eqref{onlyyou} in general is a sum of three
terms
\eq{
               V=V_F+V_D+V_{\mbox{\scriptsize NS-tad}} \,,
}
where $V_F$
is  the F-term scalar potential derived from the tree-level K\"ahler
potential and the generalized flux-induced GVW superpotential.
$V_D$ is a D-term potential related to the abelian gauge fields
arising from the dimensional reduction of the R-R four-form on
orientifold even three-cycles of the Calabi-Yau. The last term is
the flux induced NS-NS tadpole contribution that will be cancelled
by localized sources upon invoking   R-R tadpole cancellation.
We remark that $V_F$ is the scalar potential used in the string
phenomenological
studies of \cite{Micu:2007rd,Palti:2007pm,Blumenhagen:2015kja}.

%%%%%%%%%%%%%%%%%%%%%%%%%%%%%%%%%%%%%%%%%%%%%%%
%%%%%%%%%%%%%%%%%%%%%%%%%%%%%%%%%%%%%%%%%%%%%%%
%%%%%%%%%%%%%%%%%%%%%%%%%%%%%%%%%%%%%%%%%%%%%%%
%%%%%%%%%%%%%%%%%%%%%%%%%%%%%%%%%%%%%%%%%%%%%%%
%%%%%%%%%%%%%%%%%%%%%%%%%%%%%%%%%%%%%%%%%%%%%%%
%%%%%%%%%%%%%%%%%%%%%%%%%%%%%%%%%%%%%%%%%%%%%%%
%%%%%%%%%%%%%%%%%%%%%%%%%%%%%%%%%%%%%%%%%%%%%%%
%%%%%%%%%%%%%%%%%%%%%%%%%%%%%%%%%%%%%%%%%%%%%%%

\section{Review of double field theory}
\label{sec_review}

In this section, we briefly review the salient features of DFT important for 
our subsequent discussion. For a more detailed introduction we would 
like to refer to the reviews \cite{Aldazabal:2013sca,Berman:2013eva,Hohm:2013bwa}.

%%%%%%%%%%%%%%%%%%%%%%%%%%%%%%%%%%%%%%%%%%%%%%%
%%%%%%%%%%%%%%%%%%%%%%%%%%%%%%%%%%%%%%%%%%%%%%%
%%%%%%%%%%%%%%%%%%%%%%%%%%%%%%%%%%%%%%%%%%%%%%%
%%%%%%%%%%%%%%%%%%%%%%%%%%%%%%%%%%%%%%%%%%%%%%%

\subsection{Basics of DFT}

Double Field Theory is defined on a space with a doubled number of dimensions, 
where in addition to the standard coordinates $x^i$ one
introduces winding coordinates $\tilde x_i$.
The two types of coordinates can be arranged into a doubled vector of the form $X^I=(\tilde x_i,x^i)$,
with $i=1,\ldots, D$.
One also introduces an $O(D,D)$-invariant metric as 
\eq{
 \eta_{IJ}=
    \left(\begin{matrix}  0 &  \delta^i{}_j \\
            \delta_i{}^j & 0 \end{matrix}\right)  ,
}
and combines the dynamical metric and Kalb-Ramond field $G_{ij}$ and $B_{ij}$  
into the generalized metric
\eq{
\label{genmetric}
 {\cal H}_{IJ}=
  \left(\begin{matrix}  G^{ij} &  -G^{ik}B_{kj} \\
   B_{ik}G^{kj} & G_{ij} -B_{ik}G^{kl}B_{lj} \end{matrix}\right)  .
}
For these matrices a coordinates basis with indices $I,J,\ldots$ has been chosen, however,
one can also employ a non-holonomic frame 
\eq{
     {\cal H}_{IJ}=E^A{}_I\, S_{AB}\, E^B{}_J \,,
}
distinguished by indices $A,B,\ldots$
from the beginning of the alphabet.
The diagonal matrix $S_{AB}$ is given by
\eq{
   S_{AB}= \left(\begin{matrix}  s^{ab} &  0 \\
            0 & s_{ab} \end{matrix}\right) ,
}
and  $s_{ab}$ denotes the flat $D$-dimensional Minkowski metric, 
which is related to the curved metric as $G_{ij} = e^a{}_i s_{ab}\op e^b{}_j$.
For the parametrization  of the generalized metric shown in 
\eqref{genmetric}, one  can then find that
\eq{
\label{gennonhol}
       E^A{}_I=  \left(\begin{matrix}  e_a{}^i &  -e_a{}^k\op B_{ki} \\
            0 & e^a{}_i \end{matrix}\right) .
}

An action for DFT is determined by invoking  symmetries:
first, one requires the action  to be invariant under local diffeomorphisms of the
doubled coordinates $X^I$, that is \ $(\tilde x_i,x^i)\to (\tilde x_i+\tilde\xi_i(X),x^i+\xi^i(X))$.
Second, the action should be  invariant under a global (or rigid) $O(D,D)$ symmetry.
It has been realized that for manifest $O(D,D)$ invariance and for closure of the algebra 
of infinitesimal diffeomorphisms, one has to impose the so-called strong constraint
\eq{
  \label{strong_c}
  \partial_i A\, \tilde\partial^i B +  \tilde\partial^i A\, \partial_i B=0\, ,
}
where $\tilde \partial^i$ denotes the derivative with respect to the winding coordinate $\tilde x_i$.
There exist two formulations of a DFT action, 
which differ by terms that  are either total derivatives or are
vanishing due to the strong constraint \eqref{strong_c}. 
For our purposes it is convenient to use the  so-called flux formulation of DFT,
which we review in the following.

%%%%%%%%%%%%%%%%%%%%%%%%%%%%%%%%%%%%%%%%%%%%%%%
%%%%%%%%%%%%%%%%%%%%%%%%%%%%%%%%%%%%%%%%%%%%%%%
%%%%%%%%%%%%%%%%%%%%%%%%%%%%%%%%%%%%%%%%%%%%%%%
%%%%%%%%%%%%%%%%%%%%%%%%%%%%%%%%%%%%%%%%%%%%%%%

\subsection{The flux formulation of DFT}

The flux formulation of DFT has been developed in \cite{Aldazabal:2011nj,Geissbuhler:2011mx,Grana:2012rr}
and is, as has been shown in \cite{Hohm:2010xe}, 
related to earlier work of Siegel \cite{Siegel:1993xq,Siegel:1993th}.
In a frame with flat indices, the action in the NS-NS sector is given by 
\begin{align}
  \nonumber
  &\mathcal S_{\rm NS\op NS}=\frac{1}{2\op \kappa_{10}^2}\int d^{D} X \: e^{-2d}\, \bigg[ \\[4pt]
  \nonumber
  &\hspace{32pt}\hspace{16pt}{\cal F}_{ABC} {\cal F}_{A'B'C'}\left( \frac{1}{4}\op {S}^{AA'} \eta^{BB'} \eta^{CC'} 
  -\frac{1}{12}\op {S}^{AA'} {S}^{BB'} {S}^{CC'} 
    -\frac{1}{6}\hspace{4pt} {\eta}^{AA'} \eta^{BB'} \eta^{CC'} \right)\\[4pt]
 \label{dftactionfluxb}
 &\hspace{32pt}+ {\cal F}_A \op {\cal F}_{A'} \op \Big( \eta^{AA'}-{S}^{AA'} \Big)
 \hspace{10pt}\bigg]\, , 
\end{align}
where we used $d^{D} X=d^D x \wedge d^D\tilde x$. The definition of $e^{-2d}$ contains the 
dilaton $\phi$ and the
determinant of the metric $G_{ij}$, and reads
$\exp(-2d)=\sqrt{-G} \exp(-2\phi)$. 
Throughout most parts of the upcoming computation we set
$2\kappa_{10}^2=1$ and only introduce mass scales $M_{\rm s}$ or  $M_{\rm
  Pl}$  when necessary.
The  ${\cal F}_A$ are expressed as
\eq{
  \label{res_012}
     {\cal F}_A=\Omega^B{}_{BA}+2 E_A{}^I \partial_I d \,,
}
with the generalized Weitzenb\"ock connection
\eq{
  \label{flux_002}
      \Omega_{ABC}=E_A{}^I (\partial_I E_B{}^J) E_{CJ}\, .
}
The three-index object $\mathcal F_{ABC}$ is the anti-symmetrization of $\Omega_{ABC}$, that is\op\footnote{
Our convention is that the anti-symmetrization of $n$ indices contains a factor of $1/n!$.}
\eq{
  \label{flux_001}
  \mathcal F_{ABC} = 3\, \Omega_{[\ul A \ul B \ul C]} 
  \,.
}

The Ramond-Ramond sector of  DFT
has been analyzed in 
\cite{Rocen:2010bk,Hohm:2011zr,Hohm:2011dv,Hohm:2011cp,Jeon:2012kd}.
We  note that the fields
transform in  the spinor representation of $O(10,10)$ so that one
can expand
\eq{
{\cal G}=\sum_{n} {1\over n!}\, G^{(n)}_{i_1\ldots i_n}\, e_{a_1}{}^{i_1}\ldots 
        e_{a_n}{}^{i_n}\, \Gamma^{a_1\ldots a_n}|0\rangle\, ,
}
where  $\Gamma^{a_1\ldots a_n}$ defines the totally anti-symmetrized
product of $n$ $\Gamma$-matrices.
Similarly, we combine the R-R gauge potentials $C^{(2n)}$ into
a spinor $\tilde{\mathcal C}$.
Then, as shown in \cite{Geissbuhler:2011mx}, one 
can  define  the R-R field strengths as
\eq{
    {\cal G}=\slnabla\op {\tilde{\mathcal C}} \,,
}
with the generalized fluxed Dirac operator given by
\eq{
\label{newdirac}
    \slnabla=\Gamma^A D_A -{1\over 3}\op \Gamma^A \op {\cal F}_A-{1\over 6}\op
    \Gamma^{ABC} \op {\cal F}_{ABC}\, .
}
For type IIB on a CY, the only relevant R-R form is the three-form field
strength, whose action is
\eq{
  \label{dft_action_rr}
  \mathcal S_{\rm R\op R}={1\over 2\op\kappa^2_{10}}\,\int d^{D} X \; \bigg[ 
  \hspace{1pt} -\frac{1}{12}\, {S}^{AA'} {S}^{BB'} {S}^{CC'}\,
{\cal G}_{ABC} \, {\cal G}_{A'B'C'}  \biggr] \,.
}

%%%%%%%%%%%%%%%%%%%%%%%%%%%%%%%%%%%%%%%%%%%%%%%
%%%%%%%%%%%%%%%%%%%%%%%%%%%%%%%%%%%%%%%%%%%%%%%
%%%%%%%%%%%%%%%%%%%%%%%%%%%%%%%%%%%%%%%%%%%%%%%
%%%%%%%%%%%%%%%%%%%%%%%%%%%%%%%%%%%%%%%%%%%%%%%

\subsection{Compactification}

Taking the point of view that DFT is not only defined on a toroidal background,
our aim in this paper is to study compactifications of the  DFT action
on Calabi-Yau three-folds.
In particular, we are interested in the resulting scalar potential for
the moduli of the Calabi-Yau that is generated
by background fluxes. 
The fluxes $\mathcal F_{ABC}$ and $\mathcal F_A$ are treated as small deviations from the 
Calabi-Yau background. In terms of the vielbeins \eqref{gennonhol}, this 
requirement can be expressed as follows
\eq{
\label{new}
  E^A{}_I = \mathring E^A{}_I + \ov E^A{}_I + \mathcal O\bigl(\ov E^2\bigr) \,,
  \hspace{50pt} \ov E^A{}_I \ll 1\,,
}
where $\mathring E^A{}_I$ describes the Calabi-Yau background and 
$\ov E^A{}_I$ encodes the fluxes. Using this expansion in \eqref{flux_002},  
for \eqref{res_012} and \eqref{flux_001} we obtain up to first order in $\ov E$ 
\eq{
  \mathcal F_{ABC} = \mathring{\mathcal F}_{ABC} + \ov{\mathcal F}_{ABC} 
  + \mathcal O\bigl(\ov E^2\bigr) \,, \qquad
 \mathcal F_{A} = \mathring{\mathcal F}_{A} + \ov{\mathcal F}_{A} 
  + \mathcal O\bigl(\ov E^2\bigr) \,.
}
Note that  $ \mathring{\mathcal F}_{ABC} $ and  $ \mathring{\mathcal
  F}_{A} $  are computed using only the Calabi-Yau 
generalized vielbein $\mathring E^A{}_I$.  
Since  $\mathring E^A{}_I$
satisfies the DFT equations of motion, these fluxes do not generate
a scalar potential for the moduli of the CY and will be neglected.
 We consider the action \eqref{dftactionfluxb} up to second order
in the deviations $\ov E{}^A{}_I$. Since ${\mathcal
  F}_{ABC}$ and ${\mathcal F}_{A}$ 
start at linear
order in $\ov E$, it follows that all other quantities appearing in the action 
are those
of the Calabi-Yau background.

In order to define the starting point of our investigation,
let us specify the setting considered in this paper:

\begin{itemize}

\item We limit our discussion to the internal Calabi-Yau part, and ignore the 
remaining directions. The exterior derivative $d$ is understood 
to only contain derivatives with respect to the internal coordinates,
and similarly for the Hodge-star operator $\star$.

\item For the underlying Calabi-Yau background we impose the strong
  constraint.  In particular, the metric 
of the Calabi-Yau three-fold is denoted by $g_{ij}$ and only depends
on the usual coordinates $x^i$.
It appears in \eqref{gennonhol} via the vielbeins $e_a{}^i$ and 
is in general not known for the Calabi-Yau manifold. However, 
its complex-structure and K\"ahler deformations are well understood.
Similarly, the $B$-field of the background only depends on coordinates $x^i$, and 
furthermore satisfies $dB=0$ on the Calabi-Yau manifold. 

\item For a Calabi-Yau three-fold there are no non-trivial homological one-cycles, and hence
non-trivial fluxes  with one index cannot be supported on the CY.
For the internal part of the action \eqref{dftactionfluxb} we can therefore set
\eq{
  \ov{\mathcal F}_A = 0 \,.
}  

\item The term $\mathcal F_{ABC} \mathcal F_{A'B'C'} \eta^{AA'}\eta^{BB'}\eta^{CC'}$ 
vanishes via Bianchi identities. This will be discussed in section~\ref{sec_ops} below.

\item Analogously to Scherk-Schwarz reductions, we consider constant
  expectation values for  the fluxes 
   $\ov{\mathcal F}_{IJK}=\mathring E^A{}_I \mathring E^B{}_J \mathring E^C{}_K \ov{\mathcal F}_{ABC}$
along the Calabi-Yau 
three-fold. These are related to the geometric $H$- and $F$-flux, and to the non-geometric 
$Q$- and $R$-flux as
\eq{
  \label{def_fluxes}
     \ov{\cal F}_{ijk}=H_{ijk}\, , \qquad 
     \ov{\cal F}^i{}_{jk}=F^i{}_{jk}\, , \qquad 
     \ov{\cal F}_i{}^{jk}=Q_i{}^{jk}\, , \qquad  
     \ov{\cal F}^{ijk}=R^{ijk}\, .
}
Again, in analogy to Scherk-Schwarz reductions, for these background
fluxes we do not  impose the strong constraint,
but only the quadratic constraints given by
the  Bianchi identities, which can be found  in
equation  \eqref{bianchids1}.

\end{itemize}
In the setting explained above, the relevant part of the DFT
Lagrangian \eqref{dftactionfluxb} in the NS-NS sector (restricted to a Calabi-Yau three-fold)
reduces to \cite{Blumenhagen:2013hva}
\eq{
\label{dftactionflux}
\mathcal L_{\rm NS\op NS}=e^{-2\phi} \op
\ov{\cal F}_{IJK} \ov{\cal F}_{I'J'K'}\,
\left( \frac{1}{4}\op {\cal H}^{II'} \eta^{JJ'} \eta^{KK'} 
 -\frac{1}{12} \op {\cal H}^{II'} {\cal H}^{JJ'} {\cal H}^{KK'}\right)+\ldots
}
Note that the generalized metric $\mathcal H$ and the dilaton $\phi$ are that of the Calabi-Yau background, 
which only depends on the usual coordinates $x^i$. 
As it will be shown in the remainder   of this paper, upon dimensional
reduction, it is precisely this part of the DFT action
that can be identified with the scalar potential
of a gauged supergravity theory.
We emphasize that  the computations to be
performed go through once the
quadratic Bianchi identities for the fluxes are imposed. Requiring stronger conditions, such as
the strong constraint, 
only eliminates some of these fluxes.

As explicitly shown in \cite{Blumenhagen:2013hva}, the action \eqref{dftactionflux} can be further expanded, 
for which it turns out to be convenient
to introduce the following combinations of fluxes
\eq{
\label{orbitfluxes}
\mathfrak{H}_{ijk}&={H}_{ijk}+3\op {F}^m{}_{[\underline{ij}}\, B_{m\underline{k}]}
    +3\, {Q}_{[\underline{i}}{}^{mn} B_{m\underline{j}}\op  B_{n\underline{k}]}
    +{R}^{mnp}  B_{m[\underline{i}} \op B_{n\underline{j}}\op  B_{p\underline{k}]} \,, 
\\
\mathfrak{F}^i{}_{jk}&={F}^i{}_{jk}+2\,{Q}_{[\underline{j}}{}^{mi}\op  B_{m\underline{k}]} 
    +{R}^{mni}  B_{m[\underline{j}} \op B_{n\underline{k}]} \,,
\\
\mathfrak{Q}_k{}^{ij}&={Q}_k{}^{ij}+{R}^{mij}  B_{mk} \,, 
\\
\mathfrak{R}^{ijk}&={R}^{ijk}\,.
}
Let us emphasize that for these fluxes Bianchi identities similar to 
\eqref{bianchids1} have to be satisfied, which can be checked by explicit computation.
We come back to this question below.
For the term in \eqref{dftactionflux} containing three factors of the metric,
we then find
\eq{
\label{lag_1}
\arraycolsep1pt
\begin{array}{@{}rccllcccccllcccl@{}}
\displaystyle {\cal L}_1 = - \frac{e^{-2\phi}}{12} \op\biggl(  \hspace{6pt}
&&&  
{\mathfrak{H}}_{ijk} & {\mathfrak{H}}_{i'j'k'} & g^{ii'} & g^{jj'}  & g^{kk'}  
&+&
3\op& {\mathfrak{F}}^i{}_{jk} & {\mathfrak{F}}^{i'}{}_{j'k'} & g_{ii'} & g^{jj'} & g^{kk'} \\
&
+&3\op&{\mathfrak{Q}}_i{}^{jk} & {\mathfrak{Q}}_{i'}{}^{j'k'} & g^{ii'} & g_{jj'} & g_{kk'}
&+&&
{\mathfrak{R}}^{ijk} & {\mathfrak{R}}^{i'j'k'} & g_{ii'} & g_{jj'} & g_{kk'} 
&
\hspace{6pt}\biggr) \,,
\end{array}
}
whereas for the term in \eqref{dftactionflux} with a single factor of the metric we have
\eq{
\label{lag_2}
\arraycolsep1pt
\begin{array}{@{}rccccccccr@{}}
\displaystyle {\cal L}_2=-\frac{e^{-2\phi}}{2}\op \biggl(  \hspace{6pt}
&&
{\mathfrak{F}}^m{}_{ni} & {\mathfrak{F}}^{n}{}_{mi'} & g^{ii'} 
&+&
{\mathfrak{Q}}_m{}^{ni} & {\mathfrak{Q}}_{n}{}^{mi'} & g_{ii'} \\
&-&
{\mathfrak{H}}_{mni} & {\mathfrak{Q}}_{i'}{}^{mn} & g^{ii'} 
&-&
{\mathfrak{F}}^i{}_{mn} & {\mathfrak{R}}^{mni'} & g_{ii'}
&
 \hspace{6pt}\biggr)\, .
\end{array}
}
For the R-R sector of type IIB DFT we can perform a similar analysis. 
We introduce a three-form 
in the following way 
\eq{
\label{res_011}
\mathfrak G_{ijk} =  F^{(3)}_{ijk} 
- {\mathfrak H}_{ijk}\op C^{(0)} -
3\op {\mathfrak F}^m_{\hspace{7pt}[\ul i\ul j}\op C^{(2)}_{|m| \ul k]}
+\frac{3}{2}\op{\mathfrak Q}_{[\ul i}{}^{mn} \op C^{(4)}_{\ul j\ul k]mn}
+ \frac{1}{6}\op {\mathfrak R}^{mnp} \op C^{(6)}_{mnp\op  ijk} ,
} 
where $F^{(3)}$ denotes the background three-form flux in the R-R sector. 
Up to second order in the fluxes, the Lagrangian \eqref{dft_action_rr} can then be expressed as 
\eq{
\label{lag_3}
{\cal L}_{\rm RR}=-\frac{1}{12} \, \mathfrak G_{ijk}\,  \mathfrak G_{i'j'k'}\, g^{ii'}\,   g^{jj'}  \, g^{kk'}  \,.
}
Note that again the metric appearing in \eqref{lag_3} is that of the Calabi-Yau background.

%%%%%%%%%%%%%%%%%%%%%%%%%%%%%%%%%%%%%%%%%%%%%%%
%%%%%%%%%%%%%%%%%%%%%%%%%%%%%%%%%%%%%%%%%%%%%%%
%%%%%%%%%%%%%%%%%%%%%%%%%%%%%%%%%%%%%%%%%%%%%%%
%%%%%%%%%%%%%%%%%%%%%%%%%%%%%%%%%%%%%%%%%%%%%%%
%%%%%%%%%%%%%%%%%%%%%%%%%%%%%%%%%%%%%%%%%%%%%%%
%%%%%%%%%%%%%%%%%%%%%%%%%%%%%%%%%%%%%%%%%%%%%%%
%%%%%%%%%%%%%%%%%%%%%%%%%%%%%%%%%%%%%%%%%%%%%%%
%%%%%%%%%%%%%%%%%%%%%%%%%%%%%%%%%%%%%%%%%%%%%%%

\section{DFT action on a Calabi-Yau manifold}
\label{sec_dft_on_cy}

As we discussed above, in the DFT actions \eqref{dftactionfluxb} and \eqref{dft_action_rr}
the metric $g_{ij}$ appears explicitly. Since the metric of a Calabi-Yau three-fold is in general not
known, a dimensional reduction is not straightforward. 
However, by rewriting the various terms and expressing them via
known objects, the problem becomes tractable. 
In the following, we explain this approach in simple cases with only one type of
flux turned on, and later give the general result.

%%%%%%%%%%%%%%%%%%%%%%%%%%%%%%%%%%%%%%%%%%%%%%%
%%%%%%%%%%%%%%%%%%%%%%%%%%%%%%%%%%%%%%%%%%%%%%%
%%%%%%%%%%%%%%%%%%%%%%%%%%%%%%%%%%%%%%%%%%%%%%%
%%%%%%%%%%%%%%%%%%%%%%%%%%%%%%%%%%%%%%%%%%%%%%%

\subsection{Fluxes as operators}
\label{sec_ops}

For our subsequent discussion, it will turn out to be convenient to interpret 
the geometric $H$- and $F$-flux, as well as the non-geometric 
$Q$- and $R$-flux, as operators acting on $p$-forms. 
In particular, we have \cite{Aldazabal:2006up,Villadoro:2006ia,Shelton:2006fd}
\eq{
  \renewcommand{\arraystretch}{1.2}
  \arraycolsep3pt
  \begin{array}{l@{\hspace{7pt}}c@{\hspace{12pt}}lcl}
  H\,\wedge & :& \mbox{$p$-form} &\to& \mbox{$(p+3)$-form} \,, \\
  F\,\circ & :& \mbox{$p$-form} &\to& \mbox{$(p+1)$-form} \,, \\
  Q\,\bullet & :& \mbox{$p$-form} &\to& \mbox{$(p-1)$-form} \,, \\  
  R\,\llcorner & :& \mbox{$p$-form} &\to& \mbox{$(p-3)$-form} \,.  
  \end{array}
}
Employing a local basis $\{dx^i\}$ and the contraction $\iota_i$ satisfying $\iota_i\op dx^j = \delta_i^j$, 
this mapping can be implemented by 
\eq{
  \label{flux_ops}
  \arraycolsep1pt
  \begin{array}{l@{\hspace{4pt}}c@{\hspace{3pt}}cc@{\hspace{5pt}}cccccl}
  H \,\wedge &=&  \frac{1}{3!} & H_{ijk} & dx^i &\wedge& dx^j  &\wedge& dx^k & \,, \\[6pt]
  F \,\circ &=&  \frac{1}{2!} & F^k{}_{ij} &  dx^i  &\wedge& dx^j & \wedge & \iota_k & \,,   \\[6pt]
  Q \,\bullet &=&  \frac{1}{2!} & Q_i{}^{jk} & dx^i &\wedge& \iota_j  &\wedge& \iota_k & \,,   \\[6pt]  
  R \,\llcorner &=& \frac{1}{3!} & R^{ijk} & \iota_i &\wedge& \iota_j &\wedge& \iota_k & \,.   
  \end{array}
}
Note that our convention for a $p$-form $\eta$ is such that 
$\eta=\frac{1}{p!}\op\eta_{i_1 \ldots i_p} dx^{i_1}\wedge \ldots \wedge  dx^{i_p}$.
Furthermore, the above fluxes can be combined with the exterior derivative $d$ into a so-called twisted
differential $\mathcal D$
\eq{
  \label{d_operator_01}
  \mathcal D = d - H\wedge\: - F\circ\: - Q\bullet\: - R\,\llcorner \,.
}
Requiring this operator to be nilpotent, in particular that $\mathcal D^2=0$, leads to
a set of constraints on the fluxes \eqref{def_fluxes}. 
These constraints correspond to Bianchi identities, and take the form
\eq{
\label{bianchids1}
&0={H}_{m[\underline{ab}} {F}^{m}{}_{\underline{cd}]}  \,, \\
&0={F}^{m}{}_{[\underline{bc}}  \, {F}^{d}{}_{\underline{a}]m}+ {H}_{m[\underline{ab}} \, {Q}_{\underline{c}]}{}^{md} \,, \\
&0={F}^{m}{}_{[\underline{ab}]} \, {Q}_{m}{}^{[\underline{cd}]} - 4\, {F}^{[\underline{c}}{}_{m[\underline{a}} \, 
{Q}_{\underline{b}]}{}^{\underline{d}]m} + {H}_{mab} \, {R}^{mcd}  \,, \\
&0={Q}_{m}{}^{[\underline{bc}}  \, {Q}_{d}{}^{\underline{a}]m}+ {R}^{m[\underline{ab}} \:{F}^{\underline{c}]}{}_{md}\,, \\
&0={R}^{m[\underline{ab}} \, {Q}_{m}{}^{\underline{cd}]}  \,, \\[8pt]
&0= R^{mn[\ul a} F^{\ul b]}{}_{mn} \,, \\
&0=R^{amn}H_{bmn} - F^{a}{}_{mn} Q_b{}^{mn} \,, \\
&0=Q_{[\ul a}{}^{mn} H_{\ul b]mn} \,, \\[8pt]
&0= R^{mnl}H_{mnl} \,.
}
We will further impose 
\eq{
\label{unimod}
F^i{}_{ij} =0\, , \qquad
Q_i{}^{ij} = 0 \, ,
}
which are standard in the literature \cite{Shelton:2005cf, Wecht:2007wu}. 
Moreover, on a CY three-fold  there are no homologically non-trivial one- and
five-cycles,  so that it is justified to require that all combinations leaving effectively one
free-index are trivial. 
The five first identities in the upper block were originally obtained both by T-duality
and from Jacobi identities of a flux algebra \cite{Shelton:2005cf, Wecht:2007wu}.
By virtue of \eqref{unimod} the three identities in the second block follow by
taking appropriate index contractions in the middle identities of the upper block.
We are then left with the last identity \mbox{$R^{mnl}H_{mnl}=0$}, which in turn  implies 
\mbox{$F^{l}{}_{mn} Q_l{}^{mn}=0$}. Similar results have been reported in \cite{Robbins:2007yv}.
Let us emphasize that these two identities imply that  $\mathcal F_{ABC} \mathcal
F_{A'B'C'} \eta^{AA'}\eta^{BB'}\eta^{CC'}=0$.
Furthermore, we note that in orientifold theories  \mbox{$R^{mnl}H_{mnl}=0$} is automatically
satisfied, as $H$ and $R$ are of opposite parity with respect to the
$\mathbb Z_2$ orientifold projection $\Omega_{\rm P} (-1)^{F_L}$.
This will be discussed below.

%%%%%%%%%%%%%%%%%%%%%%%%%%%%%%%%%%%%%%%%%%%%%%%
%%%%%%%%%%%%%%%%%%%%%%%%%%%%%%%%%%%%%%%%%%%%%%%
%%%%%%%%%%%%%%%%%%%%%%%%%%%%%%%%%%%%%%%%%%%%%%%
%%%%%%%%%%%%%%%%%%%%%%%%%%%%%%%%%%%%%%%%%%%%%%%

\subsection{Lessons from one type of flux}
\label{sec_onetype}

Let us now consider situations with vanishing $B$-field, and only one type of 
non-trivial flux on the Calabi-Yau manifold.
More involved cases are discussed in the subsequent sections.

%%%%%%%%%%%%%%%%%%%%%%%%%%%%%%%%%%%%%%%%%%%%%%%
%%%%%%%%%%%%%%%%%%%%%%%%%%%%%%%%%%%%%%%%%%%%%%%

\subsubsection*{Pure H-flux}

We begin by turning on only $H$-flux in the Lagrangians \eqref{lag_1} and \eqref{lag_2}. 
For the NS-NS sector $\mathcal L_{\rm NS\op NS} = \mathcal L_1 + \mathcal L_2$ we obtain
\eq{
\label{res_006}
&\star\op \mathcal L_{\rm NS\op NS} = - \frac{e^{-2\phi}}{12} \op H_{ijk} \op  H_{i'j'k'} \op  g^{ii'}  g^{jj'}  g^{kk'} 
\star 1 = 
- \frac{e^{-2\phi}}{2}\, H\wedge\star H \,, 
}
where $\star 1 = \sqrt{g}\, d^{6}x$
is a convenient way to write the six-dimensional volume form
of the Calabi-Yau manifold. Let us note that the Hodge-star operator 
for a three-form on a Calabi-Yau three-fold can be evaluated 
using special geometry, for which the explicit form of the metric is not needed. 
For the next case we follow a similar strategy.

%%%%%%%%%%%%%%%%%%%%%%%%%%%%%%%%%%%%%%%%%%%%%%%
%%%%%%%%%%%%%%%%%%%%%%%%%%%%%%%%%%%%%%%%%%%%%%%

\subsubsection*{Pure F-flux}

We now turn to a more complicated situation and consider
pure $F$-flux in the case of vanishing $B$-field.
The Lagrangians in the NS-NS sector \eqref{lag_1} and \eqref{lag_2} 
then take the form
 \eq{
 \label{res_002}
  {\cal L}_{\rm NS\op NS}
  =-{e^{-2\phi}\over 4}\Big(  
  {{F}}^i{}_{jk}\, {{F}}^{i'}{}_{j'k'}\, g_{ii'} g^{jj'} g^{kk'} +
 2 \, {{F}}^m{}_{ni}\, {{F}}^{n}{}_{mi'}\,  g^{ii'} \Big)\,.
}

%%%%%%%%%%%%%%%%%%%%%%%%%%%%%%%%%%%%%%%%%%%%%%%

\paragraph{Part 1} For the first term, we define a three-form $\Xi_3= - {\cal D} J=F\circ J$,
where $J$ denotes the K\"ahler form of the Calabi-Yau three-fold.
Using our conventions \eqref{flux_ops}, we determine the components of $\Xi_3$ as
\eq{
    \Xi_{ijk}= F^m{}_{ij}\op J_{mk} + {\rm cyclic} \,.
}
We then consider the analogue of the kinetic term for the $H$-flux and compute
\eq{
\label{res_001}
\frac{1}{2} \, \Xi_3\wedge \star\op \Xi_3=
\left[ {1\over 4}\op F^{m}{}_{ij}\op F^{m'}{}_{i'j'}\op g_{mm'}\op g^{ii'}\op g^{jj'} -
        {1\over 2}\op F^{m}{}_{ij}\op F^{m'}{}_{i'j'}\op   I^{j'}{}_m\op I^{j}{}_{m'}\op  g^{ii'} 
\right] \star 1 \op,
}
where  $I_i{}^j = J_{im} g^{mj}$ is the complex structure of the 
Calabi-Yau three-fold, which satisfies $I_i{}^mI_m{}^j = - \delta_i{}^j$.
Note that the first term in \eqref{res_001} agrees with the first term in \eqref{res_002}.
For the second term in \eqref{res_001} we switch to a
complex coordinate basis and compute
\eq{
\label{res_003}
&-\frac{1}{2}\op F^{m}{}_{ij}\op F^{m'}{}_{i'j'}\op   I^{j'}{}_m\op I^{j}{}_{m'}\op  g^{ii'} \\
&\hspace{80pt} =\left(
  F^c{}_{ab}\op  F^b{}_{\ov a c}+
  F^{\ov c}{}_{a\ov b}\op F^{\ov b}{}_{\ov a \ov c}-
  F^{\ov c}{}_{ab}\op  F^b{}_{\ov a \ov c}-
  F^c{}_{a\ov b}\op  F^{\ov b}{}_{\ov a c}
\right) g^{a\ov a}\,.
}
Next, we note that using the second Bianchi identity in \eqref{bianchids1} (for vanishing $H$- and 
$Q$-flux) as 
$F^k{}_{a\ov b}\,  F^{\ov b}{}_{\ov a  k}+{\rm cyclic}=0$, we can show that
\eq{
  \label{res_005}
  g^{a\ov a} F^c{}_{a\ov b}\,  F^{\ov b}{}_{\ov a c}=
  g^{a\ov a} F^{\ov c}{}_{a b}\,  F^{b}{}_{\ov a \ov c} \,.
}
We use this relation in \eqref{res_003} and obtain
\eq{
\label{xithree}
-\frac{1}{2}\op F^{m}{}_{ij}\op F^{m'}{}_{i'j'}\op   I^{j'}{}_m\op I^{j}{}_{m'}\op  g^{ii'} =\left(
  F^c{}_{ab}\op  F^b{}_{\ov a c}+
  F^{\ov c}{}_{a\ov b}\op F^{\ov b}{}_{\ov a \ov c}-
  2\,F^{\ov c}{}_{ab}\op  F^b{}_{\ov a \ov c}\right) g^{a\ov a} \,.
}

%%%%%%%%%%%%%%%%%%%%%%%%%%%%%%%%%%%%%%%%%%%%%%%

\paragraph{Part 2} Equation \eqref{res_001} together with \eqref{xithree} only partially
 reproduce the Lagrangian \eqref{res_002}.
Let us therefore consider a second term given by
$\Xi_6=\Omega\wedge \Xi_3=-\Omega\wedge ({\cal D} J)$,
where $\Omega$ is the holomorphic three-form of a Calabi-Yau three-fold.
In components, $\Xi_6$ reads
\eq{
  \label{res_008}
    \Xi_{ijklmn}=20\, \Omega_{[\ul{ijk}}\, \Xi_{\ul{lmn}]} 
    \,, 
}      
and the corresponding kinetic term can be evaluated using the relations shown in 
appendix~\ref{app_relation}. After a somewhat tedious but straightforward computation, we find
\eq{
\label{xisix}
-\frac{1}{2} \, \Xi_6\wedge \star\op  \ov\Xi_6 
     =-2 \left[ 
     F^{\ov c}{}_{a b}\op  F^c{}_{\ov a \ov b}\op g_{c\ov c}\op  g^{a\ov a}\op  g^{b\ov b}
     -2\op F^{\ov c}{}_{ab}\op  F^b{}_{\ov a \ov c}\op g^{a\ov a}
     \right] \star 1 \,.
}

%%%%%%%%%%%%%%%%%%%%%%%%%%%%%%%%%%%%%%%%%%%%%%%

\paragraph{Part 3} One of the terms in \eqref{xisix} has the required form to 
complete the matching with the Lagrangian \eqref{res_002}. However, also 
an additional term was generated.  
In order to cancel this new term, we consider 
$\Xi_4=-{\cal D} \Omega= F\circ \Omega$.
Note that $\Xi_4$ is a $(2,2)$-form, since there are no cohomological $(3,1)$-forms on a
Calabi-Yau three-fold.
In components, we find 
\eq{
    \Xi_{ijkl}= 6\, \Omega_{[\ul{ij}m}\,F^m{}_{\ul{kl}]} 
    \,, 
}
and the kinetic term can again be evaluated using the identities given in 
appendix~\ref{app_relation}.
In particular, we have
\eq{
\label{res_004}
\frac{1}{2} \,  \Xi_4\wedge \star\op \ov\Xi_4
=2\,  F^{\ov c}{}_{a b}\,  F^c{}_{\ov a \ov b}\, g_{c\ov c}\,  g^{a\ov a}\,  g^{b\ov b}\: \star 1 \,.
}

%%%%%%%%%%%%%%%%%%%%%%%%%%%%%%%%%%%%%%%%%%%%%%%

\paragraph{Combining the individual terms} We can now combine equation \eqref{res_001} and \eqref{xithree},
together with \eqref{xisix} and \eqref{res_004}. Since the prefactors have been 
chosen in an appropriate way, with the help of the Bianchi identity \eqref{res_005} we obtain
\eq{
\label{final}
\star {\cal L}_{\rm NS\op NS}
&=-e^{-2\phi}\left[\:
      {1\over 2}\, \Xi_3\wedge \star\op \Xi_3 +
      {1\over 2}\, \Xi_4\wedge \star\op \ov\Xi_4 -
      {1\over 2}\, \Xi_6\wedge \star\op \ov\Xi_6 \:
      \right] .
}
Substituting the definitions for $\Xi_3$, $\Xi_4$ and $\Xi_6$ given above, we arrive at
\eq{      
\label{res_007}
\star {\cal L}_{\rm NS\op NS}
=-e^{-2\phi}\,\biggl[ \hspace{14pt} &
  \frac{1}{2}\, (F\circ J)\wedge \star (F\circ J) +
  \frac{1}{2}\, (F\circ \Omega)\wedge \star (F\circ\ov\Omega) 
\\
-&\frac{1}{2}\, (\Omega\wedge F\circ J)\wedge \star (\ov\Omega\wedge F\circ J)
\hspace{13pt}\biggr] \,.
}
Let us emphasize that this expression does not contain the metric 
of the Calabi-Yau manifold explicitly but depends only on $J$ and $\Omega$. 
It can therefore be evaluated using special geometry. We will come back to this point in  
section~\ref{sec_reduce}.

%%%%%%%%%%%%%%%%%%%%%%%%%%%%%%%%%%%%%%%%%%%%%%%
%%%%%%%%%%%%%%%%%%%%%%%%%%%%%%%%%%%%%%%%%%%%%%%

\subsubsection*{Pure Q-flux}

The case of pure $Q$-flux is similar to the situation with pure $F$-flux. 
We do not present a detailed derivation here, but only want to mention
one important technical step in the computation. In particular, 
since  there are no one-forms  on a Calabi-Yau manifold, we
have $Q\bullet J=0$ (in cohomology). This implies that 
\eq{
\label{resQJ2}
\bigl(Q\bullet\tfrac{1}{2} J^2\bigr)_{ijk}= Q_i{}^{mn}\op J_{jm}\op J_{kn}+{\rm cyclic}\,.
}
With all other fluxes set to zero the only Bianchi identity in \eqref{bianchids1} that survives is
${Q}_{m}{}^{[\underline{ij}}  \, {Q}_{k}{}^{\underline{l}]m}=0$. In a complex basis it implies
in particular
\eq{
  \label{BIQQ}
  g_{a\ov a} Q_{\ov b}^{\  a c}\,  Q_{c}{}^{\ \ov a \ov b}=
  g_{a\ov a} Q_{b}^{\ a  \ov c }\,  Q_{\ov c}{}^{\ \ov a b} 
 \,.
}
Using this identity, the relation \eqref{resQJ2}, as well as 
properties shown in appendix \ref{app_relation},  we then obtain
\eq{    
\label{res_009}  
\star {\cal L}_{\rm NS\op NS}
=-e^{-2\phi}\,\biggl[ \hspace{13pt} &
  \frac{1}{2}\, (Q\bullet \tfrac{1}{2}J^2)\wedge \star (Q\bullet \tfrac{1}{2}J^2) 
  +
  \frac{1}{2}\, (Q\bullet \Omega)\wedge \star (Q\bullet\ov\Omega) 
\\
-&\frac{1}{2}\, (\Omega\wedge Q\bullet \tfrac{1}{2}J^2)\wedge \star (\ov\Omega\wedge 
Q\bullet \tfrac{1}{2}J^2)
\hspace{12pt}\biggr] \,.
}
Note that this expression  is completely analogous to the pure $F$-flux result \eqref{res_007}.

%%%%%%%%%%%%%%%%%%%%%%%%%%%%%%%%%%%%%%%%%%%%%%%
%%%%%%%%%%%%%%%%%%%%%%%%%%%%%%%%%%%%%%%%%%%%%%%

\subsubsection*{Pure R-flux}

Finally, we analyze the case of pure $R$-flux. To do so, we first note that the volume
form of a Calabi-Yau three-fold can  be expressed using the K\"ahler form $J$ as
\eq{
  \sqrt{g} \, d^6 x = \frac{1}{6!} \, \epsilon_{i_1 \ldots i_6} \,dx^{i_1} \wedge \ldots
  \wedge dx^{i_6}  = \frac{1}{3!} \op J^3 \,,
}
where $\epsilon_{i_1 \ldots i_6}$ is the totally anti-symmetric tensor with $\epsilon_{123456}=\sqrt{g}$.
Recalling the definition of $R \,\llcorner$ shown in \eqref{flux_ops}, we can compute
\eq{
  R \,\llcorner \left( \frac{1}{3!} \op J^3 \right) = - \frac{1}{3!\,3!}\op  R^{ijk}\op \epsilon_{ijk pqr}
  dx^p\wedge dx^q \wedge dx^r \,.
}
Next, with the relation $\epsilon^{m_1 m_2 m_3 i_1 i_2 i_3} \epsilon_{m_1 m_2 m_3 j_1 j_2 j_3}
= 3!\, 3! \,\delta_{j_1}^{[\ul{i_1}}\, \delta_{j_2}^{\ul{i_2}} \, \delta_{j_3}^{\ul{i_3}]}$ (valid for a manifold with Euclidean
signature), we compute
\eq{
  R \,\llcorner \left( \tfrac{1}{3!} \op J^3 \right) \wedge \star R \,\llcorner \left( \tfrac{1}{3!} \op J^3 \right)
  = \frac{1}{3!} \, R^{ijk} \op  R^{i'j'k'} \op  g_{ii'}  g_{jj'}  g_{kk'} \star 1 \,.
}
For the Lagrangians \eqref{lag_1} and \eqref{lag_2}  we therefore obtain
\eq{
\label{res_010}
\star\op \mathcal L_{\rm NS\op NS} = 
- \frac{e^{-2\phi}}{2}\,  R \,\llcorner \left( \tfrac{1}{3!} \op J^3 \right) \wedge \star R \,\llcorner \left( \tfrac{1}{3!} \op J^3 \right) \,.
}
For future use we also note the following relation, which can be verified using for instance 
equation \eqref{res_037}:
\eq{
\label{resOR}
\bigl(R \,\llcorner  \Omega \bigr)\wedge \star \bigl(R \,\llcorner  \ov\Omega\op\bigr) -
\bigl(\Omega \wedge R \,\llcorner  \tfrac{1}{3!} \op J^3 \bigr)
 \wedge \star \bigl(\ov\Omega\wedge R \,\llcorner  \tfrac{1}{3!}J^3 \bigr) =0 \, .
}

%%%%%%%%%%%%%%%%%%%%%%%%%%%%%%%%%%%%%%%%%%%%%%%
%%%%%%%%%%%%%%%%%%%%%%%%%%%%%%%%%%%%%%%%%%%%%%%
%%%%%%%%%%%%%%%%%%%%%%%%%%%%%%%%%%%%%%%%%%%%%%%
%%%%%%%%%%%%%%%%%%%%%%%%%%%%%%%%%%%%%%%%%%%%%%%

\subsection{General result}

In the last section we have shown how the DFT Lagrangians \eqref{lag_1} and \eqref{lag_2} 
can be rewritten, such that only the K\"ahler form $J$ and the 
holomorphic three-form $\Omega$ appear explicitly.
We assumed a vanishing $B$-field, and have considered only one type of 
flux being present.
In this section, we now allow for all types of fluxes $H$, $F$, $Q$ and $R$ being present 
simultaneously (subject to Bianchi identities).

Motivated by our previous results, 
summarized in equations \eqref{res_006}, \eqref{res_007}, \eqref{res_009}, \eqref{res_010} and \eqref{resOR},
we define the three-form 
\eq{
   \chi=
   -H-F\circ (iJ)-Q\bullet \left({(iJ)^2\over 2!}\right) -R \,\llcorner
     \left({(iJ)^3\over 3!}\right) .
}
Noting then that the K\"ahler form $J$ is closed under $d$, we can write 
\eq{
  \label{def_chi}
    \chi &={\cal D} \left(e^{iJ}\right) .
}
Similarly, recalling that the holomorphic three-form is $d$-closed,
we define a multi-form of even degree as
\eq{
  \label{def_psi}
   \Psi=-H\wedge \Omega-F\circ \Omega-Q\bullet \Omega  -R \, \llcorner \Omega
   ={\cal D}\,\Omega \,.
} 
We  now propose the following form of the DFT Lagrangian in the NS-NS sector on
a Calabi-Yau three-fold with vanishing $B$-field
\eq{
\label{final3a}
  \star \mathcal L_{\rm NS\op NS} = -e^{-2\phi}\bigg[\hspace{18pt}
  &
  {1\over 2}\, \chi \wedge \star\op \ov\chi \;+\;{1\over 2}\op \Psi\wedge \star\op \ov\Psi \\
  -& {1\over 4}\op \big(\Omega\wedge \chi\big)\wedge\star \op \big(\ov\Omega\wedge \ov \chi\big)
  -{1\over 4}\op \big(\Omega\wedge \ov\chi\big)\wedge\star\op \big(\ov\Omega\wedge  \chi\big)
  \hspace{8pt}\biggr]\,.
}
In appendix \ref{app_proof}, we show that this Lagrangian indeed corresponds to the
full Lagrangian \eqref{dftactionflux} (for vanishing $B$-field).

%%%%%%%%%%%%%%%%%%%%%%%%%%%%%%%%%%%%%%%%%%%%%%%
%%%%%%%%%%%%%%%%%%%%%%%%%%%%%%%%%%%%%%%%%%%%%%%
%%%%%%%%%%%%%%%%%%%%%%%%%%%%%%%%%%%%%%%%%%%%%%%
%%%%%%%%%%%%%%%%%%%%%%%%%%%%%%%%%%%%%%%%%%%%%%%

\subsection{Including the $B$-field}

Let us now include a non-vanishing $B$-field, which however satisfies $dB=0$ on the Calabi-Yau 
manifold. 
The computations are completely analogous to section~\ref{sec_onetype}, provided we substitute
\eq{
  H \to \mathfrak H \,,\hspace{40pt}
  F \to \mathfrak F \,,\hspace{40pt}
  Q \to \mathfrak Q \,,\hspace{40pt}
  R \to \mathfrak R \,,     
}
where the flux orbits have been defined in \eqref{orbitfluxes}. This implies that
the Lagrangian \eqref{final3a} is the correct expression even with $B$-field, but with the twisted differential 
$\mathcal D$ in \eqref{def_chi} and \eqref{def_psi} replaced by
\eq{
  \mathcal D \quad\to\quad \mathfrak D = d - \mathfrak H\wedge\: - 
  \mathfrak F\circ\: - \mathfrak Q\bullet\: - \mathfrak R\,\llcorner \,.
}
Next, we note that using the (local) form of the flux operators \eqref{flux_ops}, we can
check that $\mathfrak D$ can be expressed in terms of $\mathcal D$ as
\eq{
  \label{res_013}
  \mathfrak D = e^{-B} \op \mathcal D \,e^{B} 
  -\frac{1}{2} \Bigl( \mathfrak Q_i{}^{mn} B_{mn} \op dx^i + \mathfrak R^{imn} B_{mn} \op  \iota_i \Bigr)
  \,.
}
On a Calabi-Yau manifold, the last two terms may be locally defined, but not globally.
This is due to the absence of non-trivial one-forms (in cohomology), and therefore we can discard them in
the following. However, in general these terms combine with the fluxes \eqref{res_012} 
into new flux orbits, analogous to \eqref{orbitfluxes}.

Employing then the relation \eqref{res_013} on a Calabi-Yau manifold, 
we can conclude that the rewritten Lagrangian for non-vanishing
$B$-field is also given by \eqref{final3a}, that is 
\eq{
\label{final3}
  \star \mathcal L_{\rm NS\op NS} = -e^{-2\phi}\bigg[\hspace{18pt}
  &
  {1\over 2}\, \chi \wedge \star\op \ov\chi \;+\;{1\over 2}\op \Psi\wedge \star\op \ov\Psi \\
  -& {1\over 4}\op \big(\Omega\wedge \chi\big)\wedge\star \op \big(\ov\Omega\wedge \ov \chi\big)
  -{1\over 4}\op \big(\Omega\wedge \ov\chi\big)\wedge\star\op \big(\ov\Omega\wedge  \chi\big)
  \hspace{8pt}\biggr]\,,
}
together with
\eq{  
  \label{res_014}
\chi&=-{\mathfrak H}-{\mathfrak F}\circ (iJ)-{\mathfrak Q}\bullet
   \left({(iJ)\wedge (iJ)\over 2}\right) -{\mathfrak R}\llcorner
   \left({(iJ)\wedge (iJ)\wedge (iJ)\over 6}\right)  \\
   &= {\mathfrak D} \op  e^{iJ} \\[6pt]
&=  e^{-B} \op \mathcal D \left( e^{B+i J} \right)\,,
}
and
\eq{
\label{res_0161}
\Psi&=-{\mathfrak H}\wedge \Omega-{\mathfrak F}\circ \Omega-{\mathfrak Q}\bullet \Omega  
  -{\mathfrak R}\llcorner  \Omega \\
  &=  \mathfrak D\op \Omega\\
  &= e^{-B} \,\mathcal D  \left( e^{B} \, \Omega\right) .
  }

%%%%%%%%%%%%%%%%%%%%%%%%%%%%%%%%%%%%%%%%%%%%%%%
%%%%%%%%%%%%%%%%%%%%%%%%%%%%%%%%%%%%%%%%%%%%%%%
%%%%%%%%%%%%%%%%%%%%%%%%%%%%%%%%%%%%%%%%%%%%%%%
%%%%%%%%%%%%%%%%%%%%%%%%%%%%%%%%%%%%%%%%%%%%%%%

\subsection{The Ramond-Ramond sector}

The R-R sector \eqref{lag_3} of the DFT action is much simpler to rewrite. Let us first 
introduce an even multi-form of R-R potentials $C^{(2n)}$ as
\eq{
  \label{res_024}
  \mathcal C = C^{(0)}+C^{(2)}+C^{(4)}+C^{(6)}+C^{(8)}+C^{(10)} \,.
}
The individual components are not all independent, but are subject to duality relations. 
Furthermore, our convention is that the forms $C^{(2n)}$ are closed on the Calabi-Yau 
three-fold, and the only non-vanishing
flux is $F^{(3)}$, corresponding to the R-R two-form.
With the help of the  operators \eqref{flux_ops}, we can express the 
flux shown in \eqref{res_011} as
\eq{
  \label{res_020}
  \mathfrak G&=F^{(3)} 
  - {\mathfrak H}\wedge C^{(0)}
  - {\mathfrak F}\circ C^{(2)}
  -{\mathfrak Q}\bullet C^{(4)}  
  -{\mathfrak R}\, \llcorner  C^{(6)}
  \\
  &= F^{(3)} + \mathfrak D \, \mathcal C
  \\
  &= F^{(3)} + e^{-B}\, \mathcal D \left( e^{B} \,\mathcal C \right) .
}
The DFT action in the Ramond-Ramond sector \eqref{lag_3} can then be written as
\eq{
\label{actioRR}
 \star \mathcal L_{\rm RR} =-\frac{1}{2}\, \mathfrak G \wedge \star \mathfrak G  \,.
}

%%%%%%%%%%%%%%%%%%%%%%%%%%%%%%%%%%%%%%%%%%%%%%%
%%%%%%%%%%%%%%%%%%%%%%%%%%%%%%%%%%%%%%%%%%%%%%%
%%%%%%%%%%%%%%%%%%%%%%%%%%%%%%%%%%%%%%%%%%%%%%%
%%%%%%%%%%%%%%%%%%%%%%%%%%%%%%%%%%%%%%%%%%%%%%%
%%%%%%%%%%%%%%%%%%%%%%%%%%%%%%%%%%%%%%%%%%%%%%%
%%%%%%%%%%%%%%%%%%%%%%%%%%%%%%%%%%%%%%%%%%%%%%%
%%%%%%%%%%%%%%%%%%%%%%%%%%%%%%%%%%%%%%%%%%%%%%%
%%%%%%%%%%%%%%%%%%%%%%%%%%%%%%%%%%%%%%%%%%%%%%%

\section{Relation to N=2 gauged supergravity}
\label{sec_reduce}

In this section, we evaluate the 
DFT actions \eqref{final3} and \eqref{actioRR} on a Calabi-Yau three-fold.
We show that the resulting scalar potential in four dimensions
is that of N=2 gauged supergravity.
As we have emphasized before, the rewritten DFT actions \eqref{final3} and \eqref{actioRR}
no longer depend on the metric explicitly, 
but only on the K\"ahler form $J$ and the holomorphic three-form $\Omega$. 
We can therefore employ special geometry to 
carry out the dimensional reduction. 

%%%%%%%%%%%%%%%%%%%%%%%%%%%%%%%%%%%%%%%%%%%%%%%
%%%%%%%%%%%%%%%%%%%%%%%%%%%%%%%%%%%%%%%%%%%%%%%
%%%%%%%%%%%%%%%%%%%%%%%%%%%%%%%%%%%%%%%%%%%%%%%
%%%%%%%%%%%%%%%%%%%%%%%%%%%%%%%%%%%%%%%%%%%%%%%

\subsection{Generalities}
\label{sec_conv_01}

Let us first introduce some notation, and recall relations in
special geometry. For more details and derivations, we would like to refer the reader
for instance to \cite{Blumenhagen:2013fgp}.

%%%%%%%%%%%%%%%%%%%%%%%%%%%%%%%%%%%%%%%%%%%%%%%
%%%%%%%%%%%%%%%%%%%%%%%%%%%%%%%%%%%%%%%%%%%%%%%

\subsubsection*{Odd cohomology} 

In the following, we consider 
a Calabi-Yau three-fold $\mathcal X$, and denote a symplectic basis for the third cohomology by
\eq{
  \label{basis_001}
  \{\alpha_{\Lambda},\beta^{\Lambda}\} \in H^3(\mathcal X) \,, \hspace{60pt}
  \Lambda =0,\ldots, h^{2,1} \,.
}
This basis can be chosen such that the only non-vanishing pairings satisfy
\eq{
  \label{symp_01}
  \int_{\mathcal X} \alpha_{\Lambda}\wedge \beta^{\Sigma} = \delta_{\Lambda}{}^{\Sigma} \,.
}
The holomorphic three-form $\Omega$  can be expanded in the basis \eqref{basis_001} as
\eq{
\label{hol_three}
\Omega=X^\Lambda\, \alpha_\Lambda - F_\Lambda\, \beta^\Lambda,
} 
where the periods $X^\Lambda$ and $F_\Lambda$ are functions of the complex-structure moduli 
${\cal U}^m$, with $i=m,\dots , h^{2,1}$. The periods $F_{\Lambda}$ can be 
determined from a holomorphic prepotential $F$ as $F_{\Lambda} = \partial F/\partial X^{\Lambda}$,
and using  $F_{\Lambda\Sigma}=\partial F_{\Lambda}/\partial X^{\Sigma} $ 
we define the so-called period matrix as
\eq{
\label{pm}
{\cal N}_{\Lambda\Sigma}=\overline{F}_{\Lambda\Sigma}+2i \, \frac{
{\rm Im}(F_{\Lambda\Gamma}) X^\Gamma \, {\rm Im}(F_{\Sigma\Delta}) X^\Delta}{
           X^\Gamma \,{\rm Im}(F_{\Gamma\Delta}) X^\Delta}  \,.
}
This matrix can  be used to determine
\eq{
\label{hodgeperiod}
\int_{\mathcal X}  \alpha_{\Lambda} \wedge \star \op \alpha_{\Sigma}
   &=- \bigl({\rm Im}\,{\cal N}\op\bigr)_{\Lambda\Sigma}
   -\left[ \bigl({\rm Re}\,{\cal N}\op\bigr)\bigl({\rm Im}\,{\cal N}\op\bigr)^{-1} 
   \bigl({\rm Re}\,{\cal N}\op\bigr)\right]_{\Lambda\Sigma} \,, \\[8pt]
\int_{\mathcal X}   \alpha_{\Lambda} \wedge \star\op \beta^{\Sigma}
   &=- \left[ \bigl({\rm Re}\,{\cal N}\op\bigr)\bigl({\rm Im}\,{\cal N}\op\bigr)^{-1}\right]
   _{\Lambda}^{\hspace{8pt}\Sigma}\,, \\[8pt]
\int_{\mathcal X}  \beta^{\Lambda} \wedge\star\op  \beta^{\Sigma}
  &=- \left[ \bigl({\rm Im}\,{\cal N}\op\bigr)^{-1} \right]^{\Lambda\Sigma}\,.
}
For later convenience, 
we also define 
\eq{
\label{res_016}
{\cal M}_1
&=\left(\begin{matrix} 
   \mathds 1 & {\rm Re}\,{\cal N} \\
   0 & \mathds 1 \end{matrix}\right)
\left(\begin{matrix} -{\rm Im}\,{\cal N} & 0 \\
 0 &   -{\rm Im}\,{\cal N}^{-1} \end{matrix}\right)
\left(\begin{matrix} \mathds 1 & 0 \\
{\rm Re}\,{\cal N} & \mathds 1 \end{matrix}\right) \\[4pt]
&=\int_{\mathcal X} \left( \begin{matrix} 
\alpha_{\Lambda} \wedge\star\op\alpha_{\Sigma} & 
\alpha_{\Lambda} \wedge\star\op\beta^{\Sigma} \\[4pt] 
\beta^{\Lambda} \wedge\star\op\alpha_{\Sigma} & 
\beta^{\Lambda} \wedge\star\op\beta^{\Sigma} 
\end{matrix}
\right).
}

%%%%%%%%%%%%%%%%%%%%%%%%%%%%%%%%%%%%%%%%%%%%%%%
%%%%%%%%%%%%%%%%%%%%%%%%%%%%%%%%%%%%%%%%%%%%%%%

\subsubsection*{Even cohomology}

For the $(1,1)$- and $(2,2)$-cohomology of $\mathcal X$  we introduce bases of the form
\eq{
  \label{res_033a}
  \arraycolsep2pt
  \begin{array}{ccl}
  \{ \omega_{\lab A} \}  &\in&  H^{1,1}(\mathcal X) \,, \\[5pt]
  \{ \sigma^{\lab A} \} & \in & H^{2,2}(\mathcal X) \,,
  \end{array}
  \hspace{50pt}\lab A = 1,\ldots, h^{1,1} \,.
}
For later convenience, we can group these two- and four-forms together with the zero- and six-form of
the Calabi-Yau three-fold. In particular, we write
\eq{
  \label{res_033}
  \arraycolsep2pt
  \begin{array}{ccl}
  \{ \omega_{ A} \}  &=&  \displaystyle \bigl\{ \tfrac{\sqrt{g}}{{\cal V}} \op dx^6,\, \omega_{\lab A} \bigr\}\,, \\[6pt]
  \{ \sigma^{ A} \} &=& \bigl\{ 1,\,\sigma^{\lab A}\bigr\} \,,
  \end{array}
  \hspace{50pt}A = 0,\ldots, h^{1,1} \,,
}
where $\mathcal V = \int_{\mathcal X} \sqrt{g} \op d^6x$ is the volume of the 
Calabi-Yau three-fold $\mathcal X$. 
These two bases can be chosen such that
\eq{
  \int_{\mathcal X} \omega_{A}\wedge \sigma^{ B} = \delta_{ A}{}^{ B} \,.
}
The triple intersection numbers corresponding to the bases \eqref{res_033a} are given by
\eq{
  \label{res_025}
  \kappa_{\lab{ABC}} = \int_{\mathcal X} \omega_{\lab A}\wedge \omega_{\lab B}
  \wedge \omega_{\lab C} \,.
}
The K\"ahler form $J$ of the Calabi-Yau three-fold $\mathcal X$
and the Kalb-Ramond field $B$ are expanded in the basis $\{\omega_{\lab A}\}$ in the following way
\eq{
J=t^{\lab A} \op \omega_{\lab A} \,, \hspace{70pt}
B = b^{\lab A}\op\omega_{\lab A}\,,
}
which can be combined into a complex field $\mathcal J$ as
\eq{
  \mathcal J = B + i J = \bigl(b^{\lab A} + i \op t^{\lab A} \bigr)\op \omega_{\lab A}
  = \mathcal J^{\lab A} \omega_{\lab A}\,.
}

%%%%%%%%%%%%%%%%%%%%%%%%%%%%%%%%%%%%%%%%%%%%%%%
%%%%%%%%%%%%%%%%%%%%%%%%%%%%%%%%%%%%%%%%%%%%%%%

\subsubsection*{$B$-twisted Hodge-star operator and Mukai pairings}

For later convenience, let us define the so-called Mukai pairing between 
forms $\rho$ and $\nu$. It is given by
\eq{
  \label{res_036}
  \langle \rho, \nu \rangle = \bigl[ \rho\wedge \lambda(\nu) \bigr]_{\rm top} \,,
}
where the projection operator $\lambda$ acts on $2n$-forms  as  
$\lambda(\rho^{(2n)}) = (-1)^n \rho^{(2n)}$
and on $(2n-1)$-forms as
$\lambda(\rho^{(2n-1)}) = (-1)^n \rho^{(2n-1)}$.
Furthermore, we define a $B$-twisted Hodge-star operator acting
on forms $\rho$ as \cite{Jeschek:2004wy,Benmachiche:2006df,Cassani:2007pq}
\eq{
  \star_B\op  \rho= e^{+B}  \wedge \star\,\lambda \bigl( e^{-B} \rho\bigr)\,.
}
For three-forms $\alpha_{\Lambda}$, we then find for instance that
\eq{
  \bigl\langle \alpha_{\Lambda} , \star_B \op \alpha_{\Sigma} \bigr\rangle 
  =\bigl(\alpha_{\Lambda}\wedge e^{-B} \bigr)\wedge \star\op \bigl( \alpha_{\Sigma} \wedge e^{-B} \bigr)
  \,,
}
and similarly for the others. Since for three-forms on a Calabi-Yau three-fold the factor $e^{-B}$ gives no
contribution, we can  express the matrix \eqref{res_016} also in the following way
\eq{
\label{res_0162}
{\cal M}_1
&=+\int_{\mathcal X} \left( \begin{matrix} 
\langle \alpha_{\Lambda} ,\star_B\op\alpha_{\Sigma} \rangle& 
\langle \alpha_{\Lambda} ,\star_B\op\beta^{\Sigma}\rangle \\[4pt] 
\langle \beta^{\Lambda} ,\star_B\op\alpha_{\Sigma} \rangle& 
\langle \beta^{\Lambda} ,\star_B\op\beta^{\Sigma} \rangle
\end{matrix}
\right).
}
The analogue of in \eqref{res_0162} for the even co-homology takes a very similar form. In particular,
we have
\eq{
\label{res_019}
{\cal M}_2
= -\int_{\mathcal X} \left( \begin{matrix} 
\langle \omega_A ,\star_B\op\omega_B\rangle & 
\langle \omega_A ,\star_B\op\sigma^B \rangle\\[4pt] 
\langle \sigma^A ,\star_B\op\omega_B\rangle & 
\langle \sigma^A ,\star_B\op\sigma^B \rangle 
\end{matrix}
\right),
}
where for instance
\eq{
  \bigl\langle \omega_A , \star_B \op \omega_B \bigr\rangle 
  =-\bigl(\omega_A \wedge e^{-B} \bigr)\wedge \star\op \bigl( \omega_B \wedge e^{-B} \bigr)
  \,.
}
Note that both ${\cal M}_1$ and ${\cal M}_2$ are positive definite matrices.

%%%%%%%%%%%%%%%%%%%%%%%%%%%%%%%%%%%%%%%%%%%%%%%
%%%%%%%%%%%%%%%%%%%%%%%%%%%%%%%%%%%%%%%%%%%%%%%

\subsubsection*{Fluxes}

The action of fluxes on the cohomology {\em in a local basis} has been 
given in \eqref{flux_ops}. However, for a Calabi-Yau manifold this can be
made more specific. 
Similarly to \cite{Grana:2006hr},  we define
\eq{
\label{deffluxes}
\arraycolsep2pt
\begin{array}{lcl@{\hspace{1.5pt}}l@{\hspace{40pt}}lcr@{\hspace{1.5pt}}l}
{\cal D}\alpha_\Lambda &=& \phantom{-}q_{\Lambda}{}^{ A} \omega_{ A}
&+\,  f_{\Lambda \op A}\sigma^{ A}\,,
&
{\cal D}\beta^\Lambda &=& \tilde q^{\Lambda \op A} \omega_{ A}
&+ \,\tilde f^{\Lambda}{}_ { A} \sigma^{ A}\,, 
\\[8pt]
{\cal  D}\omega_{ A}&=&- \tilde f^{\Lambda}{}_{ A} \alpha_\Lambda &+  \,
f_{\Lambda  A}\beta^\Lambda\,,
&
{\cal D}\sigma^{ A} &=& \tilde q^{\Lambda\op  A} \alpha_\Lambda &-  
\,q_{\Lambda}{}^{ A} \beta^\Lambda\,.
\end{array}
}
Here, $f_{\Lambda \op \lab A}$ and $\tilde f^{\Lambda}{}_ {\lab A}$ denote the geometric
$F$-fluxes, while $q_{\Lambda}{}^{\lab A}$ and $\tilde q^{\Lambda\op \lab A}$ are the
non-geometric $Q$-fluxes.
Furthermore, we use the following convention  for the $H$- and $R$-flux
\eq{
\label{fluxzerocomp}
\arraycolsep2pt
\begin{array}{lcl@{\hspace{70pt}}lcl}
f_{\Lambda \op0}&=&r_\Lambda\,, & \tilde f^{\Lambda}{}_0&=&\tilde r^\Lambda\,,\\[5pt]
q_{\Lambda}{}^0&=&h_\Lambda\,,& \tilde q^{\Lambda\op 0}&=&\tilde h^\Lambda\, .
\end{array}
}
Let us also note that the $H$-flux from section~\ref{sec_dft_on_cy} is related to the 
flux parameters as $H = -\tilde h^\Lambda \alpha_\Lambda + h_\Lambda \beta^\Lambda$.
For later convenience, we also define a $(2\op h^{2,1}+2)\times (2\op h^{1,1}+2)$ matrix
as follows
\eq{
  \label{res_015}
{\cal O}=\left(\begin{array}{@{}rr@{\hspace{2pt}}}
  -\tilde f^{\Lambda}{}_ { A} & \tilde q^{\Lambda\op  A} \\
  f_{\Lambda \op A} & -q_{\Lambda}{}^{ A}
\end{array}\right).
}

%%%%%%%%%%%%%%%%%%%%%%%%%%%%%%%%%%%%%%%%%%%%%%%
%%%%%%%%%%%%%%%%%%%%%%%%%%%%%%%%%%%%%%%%%%%%%%%
%%%%%%%%%%%%%%%%%%%%%%%%%%%%%%%%%%%%%%%%%%%%%%%
%%%%%%%%%%%%%%%%%%%%%%%%%%%%%%%%%%%%%%%%%%%%%%%

\subsection{Evaluating the action}
\label{sec_ev_ac}

Next, we evaluate the action derived in the previous
section for a Calabi-Yau three-fold. The action for the NS-NS sector has been shown in \eqref{final3}, and 
for the R-R sector in equation \eqref{actioRR}.

%%%%%%%%%%%%%%%%%%%%%%%%%%%%%%%%%%%%%%%%%%%%%%%
%%%%%%%%%%%%%%%%%%%%%%%%%%%%%%%%%%%%%%%%%%%%%%%

\subsubsection*{NS-NS sector -- part 1} 

We begin with the NS-NS sector, and focus on the three-form $\chi$ defined in 
\eqref{res_014}. First, we expand in the basis \eqref{res_033a}
\eq{
     e^{B+i J}=e^{\mathcal J} = 1+  \mathcal J^{\lab A} \omega_{\lab A}
     +\frac{1}{2} \Bigl[ \kappa_{\lab{ABC}} \mathcal J^{\lab B} \mathcal J^{\lab C} \Bigr] \sigma^{\lab A}
     +\frac{1}{6} \Bigl[ \kappa_{\lab{ABC}} \mathcal J^{\lab A} \mathcal J^{\lab B} \mathcal J^{\lab C} \Bigr] 
     \op \omega_0\,.
}
Using then the combined basis $\{\omega_A,\sigma^A\}$, we can  define a complex $(2h^{1,1}+2)$-dimensional vector $V_1$ in the following way
\eq{
  V_1 = \left( \begin{array}{c}
  \tfrac{1}{6}\op \kappa_{\lab{ABC}} \mathcal J^{\lab A} \mathcal J^{\lab B} J^{\lab C}\\
  \mathcal J^{\lab A} \\
  1 \\
   \tfrac{1}{2}\op \kappa_{\lab{ABC}} \mathcal J^{\lab B} \mathcal J^{\lab C} 
  \end{array}
  \right) ,
}
and employing matrix multiplication we observe that $(\omega_A \op\op\op \sigma^A)\cdot V_1 = e^{\mathcal J}$.
Next, we note that the last line in \eqref{deffluxes} can be expressed using \eqref{res_015} as
\eq{
  \label{res_017}
  \mathcal D \,\binom{\omega_A}{\sigma^A} = \mathcal O^T \binom{\alpha_{\Lambda}}{\beta^{\Lambda}}
  \,.
}
We then evaluate
\eq{
  \chi =  e^{-B} \op  \mathcal D \,e^{\mathcal J}
  =  e^{-B} \op\bigl( \alpha_{\Lambda}\hspace{4pt}\beta^{\Lambda}\bigr) \cdot \mathcal O \cdot V_1 \,,
}
and together with the matrix $\mathcal M_1$ given in equation \eqref{res_0162},
we  have
\eq{
\label{actioNSNSJ}
\int_{\mathcal X}  \chi \wedge \star\op \ov\chi
= V_1^T\cdot {\cal O}^T \cdot {\cal M}_1\cdot{\cal O}\cdot V_1 \,.
}

%%%%%%%%%%%%%%%%%%%%%%%%%%%%%%%%%%%%%%%%%%%%%%%
%%%%%%%%%%%%%%%%%%%%%%%%%%%%%%%%%%%%%%%%%%%%%%%

\subsubsection*{NS-NS sector -- part 2} 

A very similar route can be followed for the even multi-form $\Psi$ defined in equation
\eqref{res_0161}. We introduce a $(2\op h^{2,1}+2)$-dimensional
vector as
\eq{
  V_2 = \left(\begin{array}{@{}r@{}}X^{\Lambda} \\ -F_{\Lambda} \end{array}\right) \,,
}
and with the basis of three-forms given in \eqref{basis_001} we can write
\eq{
  \Omega = \bigl( \alpha_{\Lambda}\hspace{4pt}\beta^{\Lambda}\bigr) \cdot V_2 \,.
}
Analogous to \eqref{res_017}, we observe that using the matrix $\mathcal O$ defined 
in \eqref{res_015} we have
\eq{
  \mathcal D \, \binom{\alpha_{\Lambda}}{\beta^{\Lambda}}=  -\tilde{\mathcal O} \,
  \binom{\omega_A}{\sigma^A} \,,
  \hspace{40pt}
  \tilde{\mathcal O} = C\cdot \mathcal O\cdot C^T \,,
}
where we introduced a matrix $C$ defined as
\eq{
  \label{res_018}
    C=\left(\begin{matrix} 0 & +\mathds 1 \\ -\mathds 1 & 0\end{matrix}\right).
}
As it will be clear from the context, the dimensions of this symplectic structure are either
$(2\op h^{1,1}+2)\times(2\op h^{1,1}+2)$ or  $(2\op h^{2,1}+2)\times(2\op h^{2,1}+2)$.
We then obtain
\eq{
  \Psi  
  = e^{-B} \,\mathcal D  \left( e^{B} \, 
  \Omega\right) 
  = 
  -e^{-B} \, \bigl( \omega_A \hspace{4pt}\sigma^A\bigr) \cdot \tilde{\mathcal O}^T  \cdot V_2 \,,
}
and with the help of \eqref{res_019} we evaluate
\eq{
\label{actioNSNSom}
  \int_{\mathcal X}  \Psi \wedge \star\op \ov\Psi = 
  V_2^T \cdot \tilde{\mathcal O} \cdot \mathcal M_2 \cdot \tilde{\mathcal O}^T \cdot \ov V_2 \,.
}

%%%%%%%%%%%%%%%%%%%%%%%%%%%%%%%%%%%%%%%%%%%%%%%
%%%%%%%%%%%%%%%%%%%%%%%%%%%%%%%%%%%%%%%%%%%%%%%

\subsubsection*{NS-NS sector -- part 3}

Let us now consider the second line in the NS-NS action \eqref{final3}. We first note that 
for two six-forms $\rho_1$ and $\rho_2$ and with $\mathcal V$ the volume of the Calabi-Yau
three-fold $\mathcal X$ the following relation holds
\eq{
  \label{res_040}
  \int_{\mathcal X} \rho_1 \wedge \star\op \rho_2 =  \frac{1}{\mathcal V} \op \int_{\mathcal X} \rho_1
  \times \int_{\mathcal X} \rho_2 \,.
}
Using \eqref{symp_01} and the matrix $C$ defined in \eqref{res_018}, let us then determine for instance
\eq{
  \int_{\mathcal X} \Omega\wedge \chi = 
  V_2^T \cdot C \cdot \mathcal O\cdot V_1 \,.
}
The various other combinations are obtained analogously, and we can combine these results 
in the following way
\eq{
&\int_{\mathcal X} \Bigl[ \big(\Omega\wedge \chi\big)\wedge\star \op \big(\ov\Omega\wedge \ov \chi\big)
  + \big(\Omega\wedge \ov\chi\big)\wedge\star\op \big(\ov\Omega\wedge  \chi\big) \Bigr] \\
&\hspace{80pt}  = 
  \frac{1}{\mathcal V}\:
  V_2^T \cdot C \cdot \mathcal O\cdot \Bigl(  V_1 \times \ov V_1^T + \ov V_1 \times  V_1^T  \Bigr)
   \cdot \mathcal O^T\cdot C^T \cdot \ov V_2 \,.
}

%%%%%%%%%%%%%%%%%%%%%%%%%%%%%%%%%%%%%%%%%%%%%%%
%%%%%%%%%%%%%%%%%%%%%%%%%%%%%%%%%%%%%%%%%%%%%%%

\subsubsection*{R-R sector}

We finally turn to the Ramond-Ramond sector. The corresponding rewritten action 
is shown in equation \eqref{actioRR}.
We expand the R-R three-form flux $F^{(3)}$ in the basis \eqref{basis_001},
and define a corresponding $(2\op h^{2,1}+2)$-dimensional vector as
\eq{
  \label{res_030}
  F^{(3)} = -\tilde{\mathsf F}^{\Lambda} \op\alpha_{\Lambda} + {\mathsf F}_{\Lambda} \op \beta^{\Lambda}
  \hspace{40pt}\Rightarrow\hspace{40pt}
  \arraycolsep1.5pt
  \mathsf F^{(3)} = \left(\begin{array}{r}-\tilde{\mathsf F}^{\Lambda} \\ {\mathsf F}_{\Lambda} \end{array}
  \right)\,.
}
For the $B$-twisted R-R potentials, we expand the relevant contributions in the basis of even forms 
\eqref{res_033} as
\eq{
  e^{B} \op\mathcal C = 
  \mathsf C^{(0)}+
  \mathsf C^{(2)\lab A} \op \omega_{\lab A} +
  \mathsf C^{(4)}{}_{\lab A} \op \sigma^{\lab A} +  
  \mathsf C^{(6)}\op \omega_0 \,,
}
which defines a $(2\op h^{1,1} + 2)$-dimensional vector $\mathsf C$ 
\eq{
  \renewcommand{\arraystretch}{1.2}
  \mathsf C = \left(\begin{array}{@{}c@{}}
  \mathsf C^{(6)} \\
  \mathsf C^{(2)\lab A} \\
  \mathsf C^{(0)} \\
  \mathsf C^{(4)}{}_{\lab A} 
  \end{array}\right).
}
For the three-form flux given in equation \eqref{res_020}, we recall \eqref{res_017}
and determine
\eq{
  \mathfrak G = \bigl( \alpha_{\Lambda}\hspace{4pt}\beta^{\Lambda}\bigr) \cdot 
  \bigl( \mathsf F +\mathcal O \cdot \mathsf C \bigr) \,.
}
Employing finally the matrix $\mathcal M_1$ given in \eqref{res_016}, we arrive at
\eq{
  \int_{\mathcal X} \mathfrak G \wedge \star\op \mathfrak G = 
  \bigl( \mathsf F^T + \mathsf C^T \cdot \mathcal O^T \bigr) \cdot \mathcal M_1 \cdot
  \bigl( \mathsf F + \mathcal O \cdot \mathsf C \bigr) \,.
}

%%%%%%%%%%%%%%%%%%%%%%%%%%%%%%%%%%%%%%%%%%%%%%%
%%%%%%%%%%%%%%%%%%%%%%%%%%%%%%%%%%%%%%%%%%%%%%%

\subsubsection*{Final result} 

We can now combine the above results and obtain the scalar potential 
originating from evaluating the DFT actions \eqref{final3} and \eqref{actioRR}
on a Calabi-Yau three-fold. Including the appropriate pre-factors, 
we find from the above expressions
\eq{
  \label{res_022}
  V = &\hspace{14pt}
  \frac{1}{2} \op \bigl( \mathsf F^T + \mathsf C^T \cdot \mathcal O^T \bigr) \cdot \mathcal M_1 \cdot
  \bigl( \mathsf F + \mathcal O \cdot \mathsf C \bigr) \\
  & + \frac{e^{-2\phi}}{2} \op V_1^T\cdot {\cal O}^T \cdot {\cal M}_1\cdot{\cal O}\cdot V_1\\
  & + \frac{e^{-2\phi}}{2} \op   V_2^T \cdot \tilde{\mathcal O} \cdot \mathcal M_2 \cdot \tilde{\mathcal O}^T \cdot \ov V_2
  \\
  &-\frac{e^{-2\phi}}{4\op \mathcal V} \, 
  V_2^T \cdot C \cdot \mathcal O\cdot \Bigl(  V_1 \times \ov V_1^T + \ov V_1 \times  V_1^T  \Bigr)
   \cdot \mathcal O^T\cdot C^T \cdot \ov V_2 \,.
}
This scalar potential can be brought into the form given in equation (9) in
\cite{D'Auria:2007ay} (see also\cite{Cassani:2008rb}), which was shown to 
agree with the scalar potential of N=2 gauged supergravity.
To see that, we first rescale $V_{1,2}\to \sqrt{8\mathcal V}\,V_{1,2}$ and note that the 
potential \eqref{res_022} is multiplied by $M_{\rm s}^4$, where $M_{\rm Pl}^4= M_{\rm s}^4\, {\cal V}^2\, e^{-4\phi}$.
Introducing then $\Phi=\frac{1}{2}\op e^{-2\phi} \op {\cal V}$, we can write \eqref{res_022} as
\eq{
  V \to V'= &\hspace{14pt}
  \frac{M_{\rm Pl}^4}{8\op \Phi^2} \op \bigl( \mathsf F^T + \mathsf C^T \cdot \mathcal O^T \bigr) 
  \cdot \mathcal M_1 \cdot
  \bigl( \mathsf F + \mathcal O \cdot \mathsf C \bigr) \\
  & +\frac{2\op M_{\rm Pl}^4}{\Phi} \, V_1^T\cdot {\cal O}^T \cdot {\cal M}_1\cdot{\cal O}\cdot V_1\\
  & +\frac{2\op M_{\rm Pl}^4}{\Phi} \, V_2^T \cdot \tilde{\mathcal O} \cdot \mathcal M_2 \cdot \tilde{\mathcal O}^T \cdot \ov V_2
  \\
  &-\frac{8\op M_{\rm Pl}^4}{\Phi} \,
  V_2^T \cdot C \cdot \mathcal O\cdot \Bigl(  V_1 \times \ov V_1^T + \ov V_1 \times  V_1^T  \Bigr)
   \cdot \mathcal O^T\cdot C^T \cdot \ov V_2 \,.
}
Thus, we have succeeded in relating DFT on Calabi-Yau three-folds to the
scalar potential of N=2 gauged supergravity. This is a quite satisfying result in
that  DFT  not only provides the higher dimensional origin of N=4,
but also of N=2 gauged supergravity.

%%%%%%%%%%%%%%%%%%%%%%%%%%%%%%%%%%%%%%%%%%%%%%%
%%%%%%%%%%%%%%%%%%%%%%%%%%%%%%%%%%%%%%%%%%%%%%%
%%%%%%%%%%%%%%%%%%%%%%%%%%%%%%%%%%%%%%%%%%%%%%%
%%%%%%%%%%%%%%%%%%%%%%%%%%%%%%%%%%%%%%%%%%%%%%%
%%%%%%%%%%%%%%%%%%%%%%%%%%%%%%%%%%%%%%%%%%%%%%%
%%%%%%%%%%%%%%%%%%%%%%%%%%%%%%%%%%%%%%%%%%%%%%%
%%%%%%%%%%%%%%%%%%%%%%%%%%%%%%%%%%%%%%%%%%%%%%%
%%%%%%%%%%%%%%%%%%%%%%%%%%%%%%%%%%%%%%%%%%%%%%%

\section{Relation to type IIB orientifolds}
\label{sec_orient}

In this section, we show how the scalar potential \eqref{res_022} can be expressed
within the N=1 supergravity framework. Since in section~\ref{sec_reduce}
the four-dimensional theory preserved N=2 supersymmetry, 
we therefore perform an orientifold projection.
We choose this projection such that it leads to orientifold three- and seven-planes.
Related computations have appeared for instance in 
\cite{Grana:2005ny,Benmachiche:2006df,Grana:2006hr,D'Auria:2007ay}.
For completeness, here we present the full derivation of the scalar F- and D-term potentials,
and provide explicit expressions for the case of type IIB orientifolds.

%%%%%%%%%%%%%%%%%%%%%%%%%%%%%%%%%%%%%%%%%%%%%%%
%%%%%%%%%%%%%%%%%%%%%%%%%%%%%%%%%%%%%%%%%%%%%%%
%%%%%%%%%%%%%%%%%%%%%%%%%%%%%%%%%%%%%%%%%%%%%%%
%%%%%%%%%%%%%%%%%%%%%%%%%%%%%%%%%%%%%%%%%%%%%%%

\subsection{Generalities}

We begin our discussion by introducing the notation and conventions to be employed below,
and by recalling some well-known properties of type IIB orientifold compactifications
on Calabi-Yau three-folds \cite{Grimm:2004uq}.

%%%%%%%%%%%%%%%%%%%%%%%%%%%%%%%%%%%%%%%%%%%%%%%
%%%%%%%%%%%%%%%%%%%%%%%%%%%%%%%%%%%%%%%%%%%%%%%

\subsubsection*{Cohomology}

The orientifold projection we perform is 
$\Omega_{\rm P} (-1)^{F_{\rm L}} \sigma$, where $\Omega_{\rm P}$ denotes the world-sheet 
parity operator and $F_{\rm L}$ is the left-moving fermion number. 
The holomorphic involution $\sigma:{\cal X}\to {\cal X}$ acts on the 
K\"ahler form $J$ and the holomorphic $(3,0)$-form $\Omega$ of the Calabi-Yau three-fold $\mathcal X$ as
\eq{
  \label{op_01}
  \sigma^*: J\to J\,,\hspace{50pt} \sigma^*:\Omega\to-\Omega\,,
}
and the fixed loci of this involution correspond to O$7$- and O$3$-planes.
This holomorphic involution splits the cohomology into even and odd parts. This means in particular that 
\eq{
 H^{p,q}(\mathcal X) = H^{p,q}_+(\mathcal X) \oplus H^{p,q}_-(\mathcal X) \,,
 \hspace{50pt}
 h^{p,q} = h^{p,q}_+ + h^{p,q}_-\,.
}
Note that constants as well as the volume form $\sqrt{g}\, d^6x$ on $\mathcal X$ are always even
under the involution. For the other bases introduced in section~\ref{sec_conv_01}
we employ the following notation
\eq{
  \label{basis_coho}
  \arraycolsep1.5pt
  \renewcommand{\arraystretch}{1.4}
  \begin{array}{rcl@{\hspace{12pt}}lcl@{\hspace{22.5pt}}rcl@{\hspace{12pt}}lcl}
  \{\omega_{\alpha}\} & \in & H^{1,1}_+(\mathcal X) & \alpha &=& 1,\ldots,h^{1,1}_+ ,&
  \{\omega_{a}\} & \in & H^{1,1}_-(\mathcal X) & a &=& 1,\ldots,h^{1,1}_-,
  \\
  \{\sigma^{\alpha}\} & \in & H^{2,2}_+(\mathcal X) & \alpha &=& 1,\ldots,h^{1,1}_+ ,&
  \{\sigma^{a}\} & \in & H^{2,2}_-(\mathcal X) & a &=& 1,\ldots,h^{1,1}_-,
  \\
  \{\alpha_{\hat\lambda},\beta^{\hat\lambda}\} & \in & H^{3}_+(\mathcal X)
  &\hat\lambda &=& 1,\ldots,h^{2,1}_+, &
  \{\alpha_{\lambda},\beta^{\lambda}\} & \in & H^{3}_-(\mathcal X)
  &\lambda &=& 0,\ldots,h^{2,1}_-.
  \end{array}
  \\[5pt]
}

%%%%%%%%%%%%%%%%%%%%%%%%%%%%%%%%%%%%%%%%%%%%%%%
%%%%%%%%%%%%%%%%%%%%%%%%%%%%%%%%%%%%%%%%%%%%%%%

\subsubsection*{Moduli}

The fields of the ten-dimensional theory transform under the combined world-sheet parity
and left-moving fermion number in the following way
\eq{
  \Omega_{\rm P}(-1)^{F_{\rm L}} = \left\{ \begin{array}{l@{\hspace{20pt}}l}
  g,\, \phi,\, C^{(0)}, \, C^{(4)} & {\rm even}\,, \\[4pt]
  B, \, C^{(2)} &{\rm odd}\,.
  \end{array}
  \right.
}
Together with \eqref{op_01}, it then follows that  the holomorphic three-form $\Omega$ is expanded 
in the odd cohomology $H^3_-(\mathcal X)$
\eq{
  \label{exp_02}
  \Omega = X^{\lambda} \alpha_{\lambda} - F_{\lambda} \op\beta^{\lambda} \,.
}
Note that the complex-structure moduli $\mathcal U^{\mu}$ with $\mu=1,\ldots,h^{2,1}_-$ are encoded
in the holomorphic three-form. 
The K\"ahler form $J$ and the components of the ten-dimensional form
fields along the six-dimensional space $\mathcal X$ can be expanded as
\eq{
  \label{expansion_01}
  J = t^{\alpha} \omega_{\alpha} \,, \hspace{34pt}
  B = b^a \omega_a\,, \hspace{34pt}
  C^{(2)} = c^a \omega_a\,, \hspace{34pt}
  C^{(4)} = \rho_{\alpha}\op \sigma^{\alpha} \,,
}  
where the components $t^{\alpha}$ of the K\"ahler form are in string frame. Quantities in Einstein frame
will be denoted by a hat, and  the transition between string and Einstein frame is achieved by
\eq{
  \hat t^{\alpha} = e^{-\phi/2} \, t^{\alpha} \,.
}
Apart from the complex structure moduli, the remaining moduli fields in the effective 
four-dimensional theory after compactification are 
the following \cite{Grimm:2004uq}
\eq{
  \label{moduli}
  \begin{array}{l@{\hspace{3.5pt}}l}
  \tau &= C^{(0)} + i\op e^{-\phi} \,, \\[10pt]
  G^a &= c^a + \tau\op b^a \,, \\[6pt] 
  T_\alpha & \displaystyle =-\frac{i}{2}\,\kappa_{\alpha\beta\gamma}\op \hat t^\beta \hat t^\gamma
   +\rho_\alpha+\frac{1}{2}\op\kappa_{\alpha a b} \op c^a b^b
  -\frac{i}{4} \op e^\phi \op \kappa_{\alpha a b} \op {G}^a (G -\ov G)^b \,,
  \end{array}
}
where $\kappa_{\alpha\beta\gamma}$ and $\kappa_{\alpha ab}$ are the triple intersection numbers
defined in \eqref{res_025}.
Using the sum of even R-R potentials $\mathcal C$ defined in \eqref{res_024},
these moduli can be encoded in a complex and even multi-form $\Phi^{\rm ev}_c$ as 
follows \cite{Benmachiche:2006df}
\eq{
  \label{def_01}
  \Phi^{\rm ev}_c &= e^{B} \op \mathcal  C + i \op e^{-\phi}\, {\rm Re}\left(e^{B+iJ}\right)  \\[4pt]
  &= \tau + G^a\op \omega_a + T_{\alpha}\op \sigma^{\alpha} \,.
}

%%%%%%%%%%%%%%%%%%%%%%%%%%%%%%%%%%%%%%%%%%%%%%%
%%%%%%%%%%%%%%%%%%%%%%%%%%%%%%%%%%%%%%%%%%%%%%%

\subsubsection*{Fluxes}

For the R-R three-form flux $F^{(3)}$ and the various geometric and non-geometric  NS-NS
fluxes, we observe the following behavior under the combined world-sheet parity and left-moving fermion-number 
transformation. In particular, we have
\eq{
  \Omega_{\rm P}(-1)^{F_{\rm L}} = \left\{ \begin{array}{l@{\hspace{20pt}}l}
  F,\, R  & {\rm even}\,, \\[4pt]
  H, \, Q,\, F^{(3)} &{\rm odd}\,.
  \end{array}
  \right.
}
Including the holomorphic involution  $\sigma$
defined in \eqref{op_01}, recalling \eqref{basis_coho},
and employing the same notation as at the end of section~\ref{sec_conv_01},
we can deduce the non-vanishing flux components as follows
\eq{
\label{op_02}
\renewcommand{\arraystretch}{1.2}
\begin{array}{l@{\hspace{6pt}}c@{\hspace{18pt}}llll}
F^{(3)}&:& && \mathsf F_{\lambda}\,, & \tilde{\mathsf F}^{\lambda} \,,\\
H&:& && h_{\lambda}\,, &  \tilde h^{\lambda} \,,\\
F&:& f_{\hat \lambda \,\alpha}\,, &  \tilde f^{\hat\lambda}{}_{\alpha}\,, & f_{\lambda \,a}\,, &  \tilde f^{\lambda}{}_{a}\,, \\
Q&:& q_{\hat\lambda}{}^a\,, & \tilde q^{\hat\lambda\,a}\,, & q_{\lambda}{}^{\alpha} \,, & \tilde q^{\lambda\,\alpha}\,, \\
R&:& r_{\hat\lambda} \,, & \tilde r^{\hat\lambda} \,.
\end{array}
}

%%%%%%%%%%%%%%%%%%%%%%%%%%%%%%%%%%%%%%%%%%%%%%%
%%%%%%%%%%%%%%%%%%%%%%%%%%%%%%%%%%%%%%%%%%%%%%%
%%%%%%%%%%%%%%%%%%%%%%%%%%%%%%%%%%%%%%%%%%%%%%%
%%%%%%%%%%%%%%%%%%%%%%%%%%%%%%%%%%%%%%%%%%%%%%%

\subsection{F-term potential}
\label{sec_fterm}

In this section, we show how after the orientifold 
projection (part of) the scalar potential \eqref{res_022} can be expressed in terms of an F-term 
potential in an N=1 supergravity language.

%%%%%%%%%%%%%%%%%%%%%%%%%%%%%%%%%%%%%%%%%%%%%%%
%%%%%%%%%%%%%%%%%%%%%%%%%%%%%%%%%%%%%%%%%%%%%%%

\subsubsection*{General form}

The K\"ahler potential for the moduli of type IIB orientifolds with O$3$- and O$7$-planes 
takes the following general form \cite{Grimm:2004uq}
\eq{
  \label{kpot}
  \mathcal K = -\log\bigl[ -i\op (\tau-\ov\tau) \bigr] - 2\log \hat{\mathcal V}
  -\log \left[ \; i \int_{\mathcal X} \Omega\wedge \ov\Omega \:
  \right] \,,
}
where $\hat{\mathcal V} = \frac{1}{3!} \,\kappa_{\alpha\beta\gamma}\op \hat t^{\alpha}\hat t^{\beta}\hat t^{\gamma}$ 
denotes the volume of the Calabi-Yau three-fold in Einstein frame.
The superpotential in the presence of 
R-R three-form flux $F^{(3)}$ and 
general NS-NS fluxes
can be written as 
\cite{Shelton:2005cf} (see also 
\cite{Berglund:2005dm,Aldazabal:2006up,Villadoro:2006ia,Shelton:2006fd,Micu:2007rd,Cassani:2007pq,Blumenhagen:2015kja}) 
\eq{
  \label{sp}
  W = \int_{\mathcal X} \Bigl( F^{(3)} + 
  \mathcal D \op \Phi^{\rm ev}_c \Bigr) \wedge \Omega\,.
}
The resulting F-term potential is expressed via the K\"ahler-covariant derivative
$D_I W= \partial_I W + \mathcal K_I \op W$, where $\partial_I$ denotes the derivative with 
respect to the scalar fields 
mentioned above and where $\mathcal K_I = \partial_I \mathcal K$. 
With $G^{I\ov J}$ the inverse of the K\"ahler metric $G_{I\ov J} = \partial_I\op \partial_{\ov J}\op  \mathcal K$, 
we have
\eq{
  \label{res_023}
  V_F = e^{\mathcal K} \left[ \, G^{I\ov J} D_IW\, D_{\ov J} \ov W - 3\op |W|^2 \,\right] \,.
}
When using the K\"ahler potential \eqref{kpot}, the scalar F-term potential can be simplified. 
For that purpose, let us split the appearing sums into a sum over complex-structure moduli $U^{\mu}$, 
and a sum over $i=\{\tau, G^a, T_{\alpha}\}$.
Employing the no-scale property of \eqref{kpot}, that is \cite{Grimm:2004uq}
\eq{
  G^{i\ov j} \,\mathcal K_i\, \mathcal K_{\ov j} =4 \,, 
}  
and defining $\mathcal K^i = G^{i\ov j}\partial_{\ov j} \,\mathcal K$, 
we obtain
\eq{
  \label{pot_01}
  V_F = e^{\mathcal K} \op \biggl[ \hspace{2pt} &
  G^{U\ov U} D_UW D_{\ov U} \ov W  \\[-2pt]
  &+ G^{i\ov j} \op \partial_i W \op \partial_{\ov j}\ov W \\
  &+ \bigl( \mathcal K^i \partial_i W \,\ov W + {\rm c.c.} \bigr) + |W|^2   \hspace{1pt} \biggr] \,.
}

%%%%%%%%%%%%%%%%%%%%%%%%%%%%%%%%%%%%%%%%%%%%%%%
%%%%%%%%%%%%%%%%%%%%%%%%%%%%%%%%%%%%%%%%%%%%%%%

\subsubsection*{Rewriting part 1}

We now consider each line in \eqref{pot_01} separately and bring them into a form
suitable for comparison with the general expression given at the end of section~\ref{sec_dft_on_cy}. 
We start with the complex-structure moduli in the first line. For ease of notation we define
\eq{
  \label{res_038}
  \mathcal A = F^{(3)} + \mathcal D\op \Phi^{\rm ev}_c 
  = \Bigl[ F^{(3)} + \mathcal D \bigl( e^{B} \op\mathcal C\bigr)
  \Bigr]  + i  \Bigl[ e^{-\phi}\,
  \mathcal D \,{\rm Re}\op \bigl(e^{B+i J}\bigr) \Bigr],
}
for the superpotential \eqref{sp}. Let us observe that the real and imaginary part of $\mathcal A$
correspond to the three-forms \eqref{res_020} and
\eqref{res_014}, respectively. In particular, taking into account \eqref{op_02} and recalling 
that five-forms on a Calabi-Yau three-fold are trivial in cohomology, we have
\eq{
  \label{res_031}
  \mathcal A = \check{\mathfrak G} + i\op e^{-\phi}\,{\rm Re}\,\check\chi  \,,
}
where the check indicates the quantities after the orientifold projection.
Using then the relations given in \eqref{rel_02} and \eqref{rel_022},
we can write for the first line in \eqref{pot_01} 
\eq{
  \label{res_12}
  e^{\mathcal K}    G^{U\ov U} D_UW D_{\ov U} \ov W 
  = \frac{e^{\phi}}{4\op \hat{\mathcal V}^2} \left[
  \int_{\mathcal X} {\mathcal A}\wedge\star \ov {\mathcal A}
  +i\int_{\mathcal X} {\mathcal A}\wedge \ov {\mathcal A} \,\right]
  -e^{\mathcal K}\left\lvert \int_{\mathcal X} {\mathcal A}\wedge \ov \Omega \,\right\rvert^2 \hspace{-2.7pt}.
}
Using \eqref{res_031}, the first term on the right-hand side of \eqref{res_12} can be written out as follows
\eq{
   \frac{e^{\phi}}{4\op \hat{\mathcal V}^2} 
  \int_{\mathcal X} {\mathcal A}\wedge\star \ov {\mathcal A}
  =  \frac{e^{\phi}}{4\op \hat{\mathcal V}^2} \left[ \,\int_{\mathcal X} \check{\mathfrak G}\wedge \star
  \op\check{\mathfrak G}
  + e^{-2\phi} \int_{\mathcal X} 
  ({\rm Re}\,\check\chi) \wedge\star\op ({\rm Re}\,\check\chi)\:
  \right] \,.
}
The second term in \eqref{res_12} contributes to various D$p$-brane tadpoles and has to be canceled by local sources. 
Employing the relation shown in equation \eqref{res_035}, we find 
\eq{
   \frac{e^{\phi}}{4\op \hat{\mathcal V}^2} \:i
  \int_{\mathcal X} {\mathcal A}\wedge \ov {\mathcal A} &= 
  +\frac{1}{2\op\hat{\mathcal V}^2} \int_{\mathcal X} F^{(3)} \wedge  \mathcal D \,{\rm Re}\op \bigl(e^{B+i J}\bigr) \\
  &= - \frac{e^{\phi}}{2\op\hat{\mathcal V}^2} \int_{\mathcal X} \Bigl[
  ({\rm Im}\,\tau) - ({\rm Im}\,G^a)\op \omega_a + ({\rm Im}\op T_{\alpha})\op \sigma^{\alpha} \Bigr]\wedge \mathcal D F^{(3)} \,.
}
The third term on the right-hand side of \eqref{res_12} will be addressed below.

%%%%%%%%%%%%%%%%%%%%%%%%%%%%%%%%%%%%%%%%%%%%%%%
%%%%%%%%%%%%%%%%%%%%%%%%%%%%%%%%%%%%%%%%%%%%%%%

\subsubsection*{Rewriting part 2}

For the second line in \eqref{pot_01} we 
recall that $\Phi^{\rm ev}_c$ in the superpotential \eqref{sp} is given by
\eqref{def_01}. We can therefore compute
\eq{
  \partial_i W 
  =\int_{\mathcal X}  \mathcal D \op \bigl(\partial_i\Phi^{\rm ev}_c\bigr)  \wedge \Omega
  =\int_{\mathcal X}  \mathcal D \,
  \scalebox{0.8}{$\left(\begin{array}{@{\hspace{2pt}}c@{\hspace{2pt}}}1 \\ \omega_a \\ \sigma^{\alpha}\end{array}\right)$}
  \wedge \Omega\,,
}
where $i=\tau, G^a,T_{\alpha}$.
Using then the relations shown in equation \eqref{res_035} of the appendix, we obtain 
\eq{
  \partial_i W 
  =\int_{\mathcal X} 
  \scalebox{0.8}{$\left(\begin{array}{@{\hspace{2pt}}c@{\hspace{2pt}}}1 \\ -\omega_a \\ \sigma^{\alpha}\end{array}\right)$}
  \wedge  \mathcal D\op \Omega
  =\scalebox{0.8}{$\left(\begin{array}{@{\hspace{1pt}}r@{\hspace{1pt}}}
  (\mathcal D\op\Omega)^0 \\ 
  -(\mathcal D\op\Omega)_a \\ 
  (\mathcal D\op\Omega)^{\alpha}
  \end{array}\right)$}
  \,,
}
where, taking into account \eqref{op_02}, we expanded $\mathcal D\op\Omega$ in the basis \eqref{res_033}
as
\eq{
  \mathcal D\op\Omega = (\mathcal D\op\Omega)^0\op\omega_0 + 
  (\mathcal D\op\Omega)_a\op \sigma^a + (\mathcal D\op\Omega)^{\alpha}\op \omega_{\alpha}\,.
}
Let us now evaluate the second line in \eqref{pot_01}. Using the 
formula for the inverse K\"ahler metric $G^{i\ov j}$ given in
\eqref{inv_met}, we obtain
\eq{
  e^{\mathcal K}  \, G^{i\ov j} \op \partial_i W \op \partial_{\ov j}\ov W
  = e^{\mathcal K}\op \frac{4\mathcal V}{e^{2\phi}} \, 
  \int_{\mathcal X}\left[ e^{-B} \mathcal D\op \Omega \right]
  \wedge \star  \left[ e^{-B} \mathcal D\op\ov\Omega\op \right] \,,
}
where $\mathcal V$ (without the hat) denotes the volume of $\mathcal X$ in string frame.
By comparing with \eqref{res_0161} and noting that in cohomology there are no five-forms on 
a Calabi-Yau three-fold, we can identify $e^{-B} \mathcal D\op \Omega = \check\Psi$.
Furthermore, for the scalar potential evaluated at a particular point in field space, we
can use the relation \eqref{res_037}. We then find that
\eq{
  e^{\mathcal K}  \, G^{i\ov j} \op \partial_i W \op \partial_{\ov j}\ov W
  = \frac{e^{-\phi}}{4\op\hat{\mathcal V}^2}
  \int_{\mathcal X} \check\Psi  \wedge \star  \ov{\check\Psi}\,.
}

%%%%%%%%%%%%%%%%%%%%%%%%%%%%%%%%%%%%%%%%%%%%%%%
%%%%%%%%%%%%%%%%%%%%%%%%%%%%%%%%%%%%%%%%%%%%%%%

\subsubsection*{Rewriting part 3}

Next, we discuss the third line in equation \eqref{pot_01}.
With the help of the K\"ahler metric computed from the K\"ahler potential \eqref{kpot},
and after a somewhat tedious but straightforward computation, we find 
\eq{
  \mathcal K^{\tau} = -(\tau-\ov\tau) \,, \hspace{30pt}
  \mathcal K^{G^a} = - (G-\ov G)^a\,, \hspace{30pt}
  \mathcal K^{T_{\alpha}} = - (T-\ov T)_{\alpha} \,,
}
where as before $\mathcal K^i = G^{i\ov j} \partial_{\ov j} \mathcal K$.
For the derivatives of $\Phi^{\rm ev}_c$ defined in \eqref{def_01} with respect to the moduli, 
we then determine
\eq{
  \mathcal K^i \partial_i \Phi^{\rm ev}_c = - \Phi^{\rm ev}_c + \ov{ \Phi}^{\rm ev}_c 
  \,.
}
Employing  the short-hand notation \eqref{res_038} for the superpotential \eqref{sp}, we find
\eq{
  \label{res_039}
  \mathcal K^i \partial_i W =   - \int_{\mathcal X} \mathcal A \wedge \Omega
  +  \int_{\mathcal X} \ov{\mathcal  A} \wedge \Omega 
  \,.
}
Coming  back to the potential \eqref{pot_01}, using \eqref{res_039}, and 
re-arranging terms, we obtain for the third line
\eq{
   e^{\mathcal K} \Bigl[\:\bigl( \mathcal K^i \partial_i W \,\ov W + {\rm c.c.} \bigr) + |W|^2 \:\Bigr]
   = e^{\mathcal K} \left\lvert \int_{\mathcal X} {\mathcal A}\wedge \ov \Omega \,\right\rvert^2 
   - e^{\mathcal K} \left\lvert \int_{\mathcal X} \bigl( {\mathcal A}-\ov{\mathcal A}\bigr)\wedge 
   \ov \Omega \,\right\rvert^2 .
}
The first term on the right-hand side will be cancelled by the last term in \eqref{res_12}.
For the second term we recall \eqref{res_031}, \eqref{res_040} and \eqref{res_037}, and  determine
\eq{
&- e^{\mathcal K} \left\lvert \int_{\mathcal X} 
\bigl( {\mathcal A}-\ov{\mathcal A}\bigr)\wedge \ov \Omega \,\right\rvert^2 \\
&\hspace{60pt}= -  \frac{e^{-\phi}}{4\op\hat{\mathcal V}^2}\int_{\mathcal X}
 \Bigl[ ({\rm Re}\,\check\chi) \wedge\Omega \Bigr] \wedge\star \op\Bigl[ ({\rm Re}\,\check\chi)\wedge\ov\Omega\Bigr]
 \\
&\hspace{60pt}= - \frac{e^{-\phi}}{8\op\hat{\mathcal V}^2} \int_{\mathcal X} \Bigl[
  \big(\Omega\wedge \check\chi\big)\wedge\star \op \big(\ov\Omega\wedge \ov{\check\chi}\big)
 +\big(\Omega\wedge \ov{\check\chi}\big)\wedge\star\op \big(\ov\Omega\wedge  \check\chi\big)
 \Bigr] \,.
}
In the last step we noted that due to \eqref{op_02} we have $({\rm Im}\,\check\chi)\in H^3_+(\mathcal X)$ 
whereas $\Omega\in H^3_-(\mathcal X)$, and therefore $\int ({\rm Im}\,\check\chi)\wedge \Omega=0$.

%%%%%%%%%%%%%%%%%%%%%%%%%%%%%%%%%%%%%%%%%%%%%%%
%%%%%%%%%%%%%%%%%%%%%%%%%%%%%%%%%%%%%%%%%%%%%%%

\subsubsection*{Combining the results}

We finally combine the individual results obtained above to obtain the 
full scalar F-term potential. In particular, we can rewrite \eqref{pot_01} as
\begin{align}
\nonumber
  V_F =\frac{M_{\rm Pl}^4\,e^{\phi}}{2\op \hat{\mathcal V}^2} \op \int_{\mathcal X} \biggl(
  &\hspace{16pt}e^{-2\phi} \op\biggl[ \hspace{5pt}  \frac{1}{2}\op({\rm Re}\,\check\chi) \wedge\star\op ({\rm Re}\,
  \check\chi)
  +\frac{1}{2}\op   \check\Psi  \wedge \star  \ov{\check\Psi} \\
\nonumber
  &\hspace{45pt}
  -\frac{1}{4}\op\big(\Omega\wedge \check\chi\big)\wedge\star \op \big(\ov\Omega\wedge \ov{\check\chi}\big)
 -\frac{1}{4}\op\big(\Omega\wedge \ov{\check\chi}\big)\wedge\star\op \big(\ov\Omega\wedge  \check\chi\big) 
 \biggr]
 \\[0.1cm]
\nonumber
  &+\frac{1}{2}\op \check{\mathfrak G}\wedge \star\op\check{\mathfrak G} \\[0.1cm]
  &-\Bigl[   ({\rm Im}\,\tau) - ({\rm Im}\,G^a)\op \omega_a + ({\rm Im}\op T_{\alpha})\op \sigma^{\alpha} \Bigr]
  \wedge \mathcal D F^{(3)} 
  \hspace{10pt}\biggr) \,.
\label{res_041}
\end{align}
Taking into account that the prefactor is proportional to $M_{\rm s}^4$,
the first two lines match with the orientifold projected DFT actions \eqref{final3} and 
\eqref{actioRR} in the NS-NS and R-R sector. Note, however, that only the real
part of $\check\chi$ appears; the imaginary part is contained in a D-term, which we discuss in the next section. 
The third line in \eqref{res_041} corresponds to tadpole terms, which have to be cancelled 
by local sources.

%%%%%%%%%%%%%%%%%%%%%%%%%%%%%%%%%%%%%%%%%%%%%%%
%%%%%%%%%%%%%%%%%%%%%%%%%%%%%%%%%%%%%%%%%%%%%%%
%%%%%%%%%%%%%%%%%%%%%%%%%%%%%%%%%%%%%%%%%%%%%%%
%%%%%%%%%%%%%%%%%%%%%%%%%%%%%%%%%%%%%%%%%%%%%%%

\subsection{D-term potential}

We now want to consider the imaginary part of $\check\chi$, which does not appear
in the scalar F-term potential \eqref{res_041}. As mentioned before, 
we have $({\rm Im}\,\check\chi)\in H^3_+(\mathcal X)$ and therefore the only  
contribution in the DFT Lagrangian \eqref{final3} relevant here comes from
\eq{
  \label{res_050}
  \star \op {\cal L}_{H^3_+}=-{1\over 2} \,e^{-2\phi}\,\op({\rm Im}\,\check\chi) \wedge\star\op ({\rm Im}\,
  \check\chi) \,.
}
Using the definition \eqref{res_014} as well as \eqref{op_02}, we can evaluate ${\rm Im}\op\check\chi$ as
\eq{
    {\rm Im}\op\check\chi= \bigl( \alpha_{\hat\lambda}\hspace{4pt}\beta^{\hat\lambda}\bigr) \cdot 
    \binom{\tilde {\mathsf D}^{\hat\lambda}}{\mathsf D_{\hat{\lambda}}}\,,
}
where we defined
\eq{
\label{res_054}
&\tilde {\mathsf D}^{\hat\lambda} = \hspace{9pt}\tilde r^{\hat\lambda}\, \big({\cal V}-\tfrac{1}{2}\op
    \kappa_{\alpha a b}\op
    t^\alpha b^a b^b\Big) + \tilde q^{{\hat\lambda} a}\, \kappa_{a\alpha  b} \op
  t^\alpha b^b - \tilde f^{\hat\lambda}{}_{\alpha}\, t^\alpha \,, \\[2pt]
&\mathsf D_{\hat{\lambda}} = -r_{\hat\lambda}\, \big({\cal V}-\tfrac{1}{2}\op \kappa_{ \alpha a b} \op
    t^\alpha b^a b^b\Big) - q_{\hat\lambda}{}^{a}\, \kappa_{a \alpha  b}\op
  t^\alpha b^b + f_{{\hat\lambda} \alpha}\, t^\alpha \,.
}
Similarly as in section~\ref{sec_ev_ac}, we can now evaluate  \eqref{res_050}. We find
\eq{
  \label{dtermdftfull}
  \star \op {\cal L}_{H^3_+}=&-{1\over 2} \,e^{-2\phi}\, \binom{\tilde {\mathsf D}}{\mathsf D}^T 
  \cdot \check{\mathcal M}_1
  \cdot  \binom{\tilde{\mathsf  D}}{\mathsf D} \\[4pt]
  =&\hspace{15pt}{1\over 2} \,e^{-2\phi}\,
  \bigg[ 
  \hspace{15pt}
  (\mathsf D_{\hat\lambda}+{\rm Re}\,{\cal N}_{{\hat\lambda} \hat\kappa} \,\tilde{\mathsf  D}^{\hat\kappa})  
  \left({\rm Im}\,{\cal N}^{-1}\right)^{{\hat\lambda}{\hat\sigma}}
  (\mathsf D_{\hat\sigma}+{\rm Re}\,{\cal N}_{{\hat\sigma} \hat\rho}\, \tilde{\mathsf  D}^{\hat\rho})  
  \\
  &\hspace{60pt}+\tilde{\mathsf  D}^{\hat\lambda}\,{\rm Im}\,{\cal N}_{{\hat\lambda}{\hat\sigma}}\, 
  \tilde{\mathsf  D}^{\hat\sigma} 
  \bigg] \,,
}
where $\mathcal M_1$ has been defined in \eqref{res_016}, and the check indicates the restriction 
to indices $\hat\lambda = 1,\ldots, h^{2,1}_+$.
Note that \eqref{dtermdftfull} corresponds to a positive semi-definite scalar potential in four dimensions.

Let us now check that this scalar potential can be understood as a
D-term from  the N=1 supergravity point of view. We will follow the discussion
first presented in \cite{Robbins:2007yv} (see also \cite{Shukla:2015bca}). 
To begin, let us recall that in the absence of a Fayet-Iliopolous term, $\xi_a=i \delta_a W/W$, the D-term of an abelian
gauge field $A^a$ in supergravity
is given by 
\eq{   
\label{dtermfield}
      D_a=i\op\sum_i (\partial_i \mathcal K) \, \delta_a \phi_i \,,
}
where $\delta_a \phi_i$ is the variation of the chiral superfield $\phi_i$
under a gauge transformation $A^a\to A^a +d\Lambda^a$, and  $\mathcal K$ denotes again the
K\"ahler potential. 
The corresponding D-term potential reads
\eq{
  \label{res_055}
  V_D = M^4_{\rm Pl} \, \Bigl[ ({\rm Re}\op f)^{-1} \Bigr]^{ab} D_a \op D_b \,,
}
with ${\rm Re}\op f_{ab}$ the real part of the gauge kinetic function for the gauge fields.
In our case, 
the gauge fields of interest originate from the R-R four-form
$C^{(4)}$ via a dimensional reduction on three-cycles of the Calabi-Yau three-fold.
Let us therefore expand
\eq{
  \label{res_052}
  C^{(4)} = A^{\hat\lambda} \alpha_{\hat\lambda} + \tilde A_{\hat\lambda} \beta^{\hat \lambda}
  + \ldots\,,
  \hspace{50pt}\hat\lambda = 1,\ldots, h^{2,1}_+ \,,
}
where the ellipsis denote terms of different degree in the internal manifold not of importance here.
The gauge transformations of $A_{\hat\lambda}$ and $\tilde A^{\hat\lambda}$ 
have their origin in a higher-dimensional gauge symmetry. In particular,  
note that the DFT Lagrangian \eqref{actioRR} is invariant under
\eq{
  \label{res_051}
  \mathcal C \to \mathcal C + \mathfrak D \Lambda \,,
}
with $\mathcal C$ the sum of even R-R potentials \eqref{res_024}, $\mathfrak D$
was defined in \eqref{res_013}, and $\Lambda$ is a sum of odd forms.
In order to obtain the gauge transformation  
$A^{\hat\lambda}\to A^{\hat\lambda}+d\Lambda^{\hat\lambda}$ and 
$\tilde A_{\hat\lambda}\to \tilde A_{\hat\lambda}-d\tilde\Lambda_{\hat\lambda}$
in four dimensions, we therefore have to choose the gauge parameter $\Lambda$ as
\eq{
e^B\op {\cal C} \;\to\; e^B \op{\cal C}+{\cal  D}\Bigl(\Lambda^{\hat\lambda} \alpha_{\hat\lambda} -\tilde
\Lambda_{\hat\lambda} \beta^{\hat\lambda}\Bigr) \,.
}
In turn, this gauge transformation implies variations of the chiral superfields $\phi\in\{\tau,G^a,T_\alpha\}$.
Indeed, using \eqref{def_01} together with \eqref{deffluxes} and \eqref{op_02} we find that
\eq{
  \label{res_053}
  \arraycolsep2pt
  \begin{array}{lclcl}
  \tau &\to& \tau &+&\bigl( r_{\hat\lambda}\op\Lambda^{\hat\lambda} -\tilde r^{\hat\lambda} \op\tilde \Lambda_{\hat\lambda} 
    \bigr)\,, \\[6pt]
  G^a &\to& G^a &+&\bigl( q_{\hat\lambda}{}^a\op\Lambda^{\hat\lambda} -\tilde q^{\hat\lambda\op a} \op
    \tilde \Lambda_{\hat\lambda}  \bigr) \,, \\[6pt]
  T_{\alpha} &\to& T_{\alpha} &+& \bigl( f_{\hat\lambda\op \alpha}\op\Lambda^{\hat\lambda} -
    \tilde f^{\hat\lambda} {}_{\alpha}\op \tilde \Lambda_{\hat\lambda}  \bigr) \,.
  \end{array}
}
Note that due to the nilpotency of ${\cal D}$, the superpotential is invariant under transformations
of the form \eqref{res_051} and thus  no Fayet-Iliopolous parameter is generated.

In order to evaluate \eqref{dtermfield}, let us also determine the derivatives of the  K\"ahler potential
\eqref{kpot} with respect to the moduli fields \eqref{moduli}. As
in the previous section, we perform the computation in
Einstein frame, and then transform the result to string frame.
We find
\eq{
    &\partial_\tau K= \frac{i e^\phi }{2 {\mathcal V}} \left(
      {\mathcal V} - \frac{1}{2}\, \kappa_{\alpha bc} t^{\alpha} b^b b^c
    \right)\\
    &\partial_{T_\alpha} K=- \frac{i e^\phi }{2 {\mathcal V}}\, t^{\alpha}\,, 
    \hspace{40pt}
    \partial_{G^a} K= \frac{i e^\phi }{2 {\mathcal V}} \kappa_{a \beta c}  t^{\beta} b^c\,.
}
Using  these results and the transformations of the moduli fields under gauge transformations
\eqref{res_053}, we can compute the D-terms \eqref{dtermfield} as follows
\eq{
\label{res_057}
&\tilde { D}^{\hat\lambda} =\frac{e^\phi }{ 2\op{\mathcal V}}\left[ 
   \hspace{9pt}\tilde r^{\hat\lambda}\, \big({\cal V}-\tfrac{1}{2}\op
    \kappa_{\alpha a b}\op
    t^\alpha b^a b^b\Big) + \tilde q^{{\hat\lambda} a}\, \kappa_{a\alpha  b} \op
  t^\alpha b^b - \tilde f^{\hat\lambda}{}_{\alpha}\, t^\alpha \right], \\[2pt]
&D_{\hat{\lambda}} =  \frac{e^\phi }{2\op{\mathcal V}}\left[ 
  -r_{\hat\lambda}\, \big({\cal V}-\tfrac{1}{2}\op \kappa_{ \alpha a b} \op
    t^\alpha b^a b^b\Big) - q_{\hat\lambda}{}^{a}\, \kappa_{a \alpha  b}\op
  t^\alpha b^b + f_{{\hat\lambda} \alpha}\, t^\alpha \right].
}
We observe that up to an overall factor, these D-terms agree with the expressions \eqref{res_054}
obtained from a reduction of the DFT action \eqref{res_050}.
We furthermore note that the Ramond-Ramond four-form potential $C^{(4)}$ is self-dual in ten 
dimensions. The two sets of gauge fields $A^{\hat\lambda}$ and $\tilde A_{\hat\lambda}$ in \eqref{res_052} 
are therefore not independent, and in the following we choose to eliminate $\tilde A_{\hat\lambda}$ in favor
of $A^{\hat\lambda}$. Also, as argued in \cite{Robbins:2007yv},
as long as the fluxes are integer-valued one can rotate them by an
${\rm Sp}(h^{2,1}_+,\mathbb Z)$ transformation into a basis where
$\tilde r^{{\hat\lambda}}=\tilde q^{{\hat\lambda} a}=\tilde f^{{\hat\lambda}}{}_{\alpha}=0$.
This implies  that the D-term $\tilde D^{\hat\lambda}$ vanishes.\footnote{In this basis the Bianchi identities 
connecting the fluxes in the D-terms are trivially satisfied. There are further 
Bianchi identities which mix the flux parameters in the superpotential \eqref{sp}
with those in $D_{\hat\lambda}$ in \eqref{res_057}.
}

Let us finally turn to the D-term potential \eqref{res_055}. The gauge kinetic function 
for the gauge fields $A^{\hat{\lambda}}$ is given by the imaginary part of the matrix \eqref{pm}
\cite{Grimm:2004uq}, properly restricted to
indices $\hat\lambda = 1,\ldots , h^{2,1}_+$
\eq{
  f_{\hat\lambda\hat\sigma} = - \frac{i}{2}\, \bar{\mathcal N}_{\hat\lambda\hat\sigma} \,.
}  
Furthermore, ${\rm Im}\,\mathcal N$ is meant to only depend on the complex structure moduli
$U^\mu$ surviving the orientifold projection.
For the D-term potential we therefore obtain
\eq{
  \label{res_056}
   V_D &= M^4_{\rm Pl} \, \Bigl[ -2\op({\rm Im}\, \mathcal N)^{-1} \Bigr]^{\hat\lambda\hat\sigma} \op
   D_{\hat\lambda} \op D_{\hat\sigma} \\[4pt]
   &= -\frac{M^4_{\rm Pl}\, e^{2\phi} }{2\op {\mathcal V}^2}
   \Bigl[ ({\rm Im}\, \mathcal N)^{-1} \Bigr]^{\hat\lambda\hat\sigma} \op
   \mathsf D_{\hat\lambda} \op \mathsf D_{\hat\sigma} \,,
}
where $\mathsf D_{\hat\lambda}$ was defined in \eqref{res_054}.
Expressing then again the Planck mass in terms of the string scale via  $M_{\rm Pl}^4= M_{\rm s}^4\, {\cal V}^2\, e^{-4\phi}$ and noting
that the potential appears as $-V$ in the Lagrangian $\mathcal L$,
we see that 
the D-term potential \eqref{res_056} agrees with the DFT result  \eqref{dtermdftfull},
after $\tilde {\mathsf D}^{\hat\lambda}$ has been set to zero.
We therefore conclude that the scalar potential resulting from the
dimensional reduction of DFT for $h^{2,1}_+>0$ also correctly
reproduces the expected  D-term potential.

%%%%%%%%%%%%%%%%%%%%%%%%%%%%%%%%%%%%%%%%%%%%%%%
%%%%%%%%%%%%%%%%%%%%%%%%%%%%%%%%%%%%%%%%%%%%%%%
%%%%%%%%%%%%%%%%%%%%%%%%%%%%%%%%%%%%%%%%%%%%%%%
%%%%%%%%%%%%%%%%%%%%%%%%%%%%%%%%%%%%%%%%%%%%%%%
%%%%%%%%%%%%%%%%%%%%%%%%%%%%%%%%%%%%%%%%%%%%%%%
%%%%%%%%%%%%%%%%%%%%%%%%%%%%%%%%%%%%%%%%%%%%%%%
%%%%%%%%%%%%%%%%%%%%%%%%%%%%%%%%%%%%%%%%%%%%%%%
%%%%%%%%%%%%%%%%%%%%%%%%%%%%%%%%%%%%%%%%%%%%%%%

\section{Conclusions}

In this paper we have performed the dimensional reduction of the
DFT action in its flux formulation on a Calabi-Yau three-fold
with non-trivial constant fluxes turned on. The main initial obstacle
that the DFT action contained explicitly the unknown metric on the  CY
could be overcome by rewriting all contributions  to the action 
in terms of the K\"ahler form, holomorphic three-form, and operations that could be further evaluated on 
the CY using
special geometry. The induced scalar potential agrees with that
of N=2 gauged supergravity. Up to additional D-terms, a further orientifold projection
to N=1 leads to the potential derived from the generalized Gukov-Vafa-Witten superpotential
containing the non-geometric fluxes. This nicely confirms the
consistency of the whole approach.

Our results put the generalized flux-induced scalar potential on firmer
grounds, thereby lending further support to its use in 
tree-level moduli stabilization applied to string
phenomenology and cosmology.
It is known that, with all types of fluxes turned on, there does not
exist a dilute flux limit so that it is not straightforward to argue
for a consistent higher dimensional uplift of the solutions
found in the four-dimensional field theory model.
However, in view of the now established DFT origin of the
four-dimensional potential, the fate of these vacua is closely related
to  the claim that DFT, though not an
effective low-energy theory,
might be a consistent truncation of full string (field) theory.

%%%%%%%%%%%%%%%%%%%%%%%%%%%%%%%%%%%%%%%%%%%%%%%
%%%%%%%%%%%%%%%%%%%%%%%%%%%%%%%%%%%%%%%%%%%%%%%

\vspace{0.5cm}

\noindent
\emph{Acknowledgments:}
We would like to thank G.~Aldazabal, D.~L\"ust and P.~Shukla for discussions, and 
we are indebted to S.~Theisen for very useful remarks. 
Moreover, we thank X.~Gao, D.~Herschmann, O.~Loaiza-Brito and P.~Shukla for earlier  collaboration
on this project.
R.B. thanks the Bethe Center for Theoretical Physics at the University Bonn for hospitality.
A.F. thanks the Alexander von Humboldt Foundation for a grant VEN/1067599 STP,
as well as the Instituto de Fisica Teorica (IFT UAM-CSIC) in Madrid for its hospitality and support via the 
Centro de Excelencia Severo Ochoa Program under Grant SEV-2012-0249.
E.P. is supported by the ERC Advanced Grant Strings and Gravity (Grant.No. 32004).

%%%%%%%%%%%%%%%%%%%%%%%%%%%%%%%%%%%%%%%%%%%%%%%
%%%%%%%%%%%%%%%%%%%%%%%%%%%%%%%%%%%%%%%%%%%%%%%
%%%%%%%%%%%%%%%%%%%%%%%%%%%%%%%%%%%%%%%%%%%%%%%
%%%%%%%%%%%%%%%%%%%%%%%%%%%%%%%%%%%%%%%%%%%%%%%
%%%%%%%%%%%%%%%%%%%%%%%%%%%%%%%%%%%%%%%%%%%%%%%
%%%%%%%%%%%%%%%%%%%%%%%%%%%%%%%%%%%%%%%%%%%%%%%
%%%%%%%%%%%%%%%%%%%%%%%%%%%%%%%%%%%%%%%%%%%%%%%
%%%%%%%%%%%%%%%%%%%%%%%%%%%%%%%%%%%%%%%%%%%%%%%
\newpage 
\appendix

\section{Useful relations on a Calabi-Yau three-fold}

In this appendix, we collect some technical relations concerning Calabi-Yau three-folds,
which are important for the computations in the main part of the paper.

%%%%%%%%%%%%%%%%%%%%%%%%%%%%%%%%%%%%%%%%%%%%%%%
%%%%%%%%%%%%%%%%%%%%%%%%%%%%%%%%%%%%%%%%%%%%%%%

\subsection{Normalization and primitivity}
\label{app_relation}

Since a Calabi-Yau manifold is a complex K\"ahler manifold, it is useful to work in
a complex basis with indices $a$ and $\ov a$.
The hermitian metric then has non-vanishing components $g_{a\ov b}$,
whereas the almost complex structure reads $I^a{}_b=i\op \delta^a{}_b$ and
\mbox{$I^{\ov a}{}_{\ov b}=-i\op\delta^{\ov a}{}_{\ov b}$}.  
The K\"ahler form $J_{ij}=g_{im} \, I^m{}_j$ in complex coordinates 
is given by $J_{a\ov b}=i \op g_{a\ov b}$. 
For the holomorphic three-form on a Calabi-Yau three-fold, 
we employ the normalization
\eq{
  \label{res_037}
  \frac{i}{8}\, \Omega\wedge \ov\Omega = \frac{1}{6}\, J^3 \,.
}
Using \eqref{res_037}, one can show the following useful relations
\eq{
   &\Omega_{abc}\, \ov{\Omega}_{\ov a\ov b \ov c}\, g^{c\ov
           c}={8}\left( g_{a\ov a}\, g_{b\ov b}-g_{a\ov b}\, g_{b\ov a}\right) , \\[4pt]
   &\Omega_{abc}\, \ov{\Omega}_{\ov a\ov b \ov c}\, g^{b\ov b}\, g^{c\ov
           c}={16}\,  g_{a\ov a} \,, \\[4pt]
   &\Omega_{abc}\, \ov{\Omega}_{\ov a\ov b \ov c}\, g^{a\ov a}\, g^{b\ov b}\, g^{c\ov
           c}={48}\,  .
} 
Since on a Calabi-Yau three-fold there are no homologically non-trivial one- and
five-cycles, we can assume that all combinations leaving effectively one
free-index are trivial. This includes e.g.
\eq{
\label{primi}
     H\wedge J=0\,,\qquad Q\bullet J=0\,, \qquad R \op\llcorner ( J\wedge J)=0 \, ,
}
as well as the conditions \eqref{unimod}.
Note that \eqref{primi} can be considered as generalized {\it primitivity} constraints
on the fluxes.

%%%%%%%%%%%%%%%%%%%%%%%%%%%%%%%%%%%%%%%%%%%%%%%
%%%%%%%%%%%%%%%%%%%%%%%%%%%%%%%%%%%%%%%%%%%%%%%

\subsection{Relations regarding complex-structure moduli}

In this section, we derive some formulas important for section~\ref{sec_fterm}.
We begin by noting that a complex basis of $(2,1)$-forms $\chi_{\mu}$ with $\mu=1,\ldots, h^{2,1}_-$ 
is given by
\eq{
  \label{res_027}
  D_{U^{\mu} } \Omega = \chi_{\mu} \,,
}
with $D_U$ the K\"ahler covariant derivative defined below \eqref{sp}. 
In a similar fashion, a basis of $(1,2)$-forms $\ov\chi_{\ov \mu}$ can be introduced. 
The K\"ahler metric for the complex-structure moduli derived from \eqref{kpot}
is expressed as
\eq{
  \label{res_028}
  G_{\mu\ov\nu} = - \frac{\int \chi_{\mu} \wedge \ov \chi_{\ov\nu}}{\int \Omega\wedge\ov \Omega}
  \,.
}  
Next, we observe that on a Calabi-Yau three-fold the holomorphic $(3,0)$-form $\Omega$ and
the $(2,1)$-forms $\chi_{\mu}$ introduced in equation \eqref{res_027}, and their complex 
conjugates form a basis of the third cohomology.
An arbitrary complex three-form $A$ can therefore be expanded in the following way
\eq{
  A = a^0 \op \Omega + a^{\mu} \op \chi_{\mu} + \tilde a^{\mu}\op \ov\chi_{\ov\mu} + \tilde a^0 \op\ov \Omega\,.
}
Using the K\"ahler metric \eqref{res_028}, the coefficients in this expansion can be determined as
\eq{
  \label{res_029}
  \arraycolsep2pt
  \begin{array}{lcl@{\hspace{40pt}}lcl}
  a^0 &=& \displaystyle + \frac{\int A \wedge \ov \Omega}{\int \Omega\wedge \ov \Omega}\,, 
  &
  a^{\mu} &=& \displaystyle -\frac{\int A\wedge \ov\chi_{\ov \nu}}{\int \Omega\wedge\ov \Omega}\, G^{\ov \nu \mu} \,, 
  \\[14pt]
  \tilde a^0 &=& \displaystyle - \frac{\int A \wedge  \Omega}{\int \Omega\wedge \ov \Omega} \,, 
  &
  \tilde a^{\ov\mu} &=& \displaystyle -\frac{\int A\wedge \chi_{ \nu}}{\int \Omega\wedge\ov \Omega}\, G^{\nu \ov\mu} \,.
  \end{array}
}
Furthermore, we note that the Hodge-star operator acting on $\Omega$ and $\chi_{\mu}$ gives
\eq{
  \star\op \Omega = -i \, \Omega\,,
  \hspace{40pt}
  \star \op\chi_{\mu} = +i\, \chi_{\mu} \,.
}
Using the above relations, for two different complex three-forms $A$ and $B$ we can then compute
\eq{
  \label{exp_01}
  \int A\wedge \star \ov B = i \int\Omega\wedge\ov\Omega \times
  \left[ a^0 \op \ov b^0 + \tilde a^0 \op\ov{\tilde b}^0 + a^{\mu} \op G_{\mu\ov\nu} \op \ov b^{\nu}
  + \ov{\tilde b}^{\mu} \op G_{\mu \ov \nu} \op \tilde a^{\ov \nu}
  \right].
}
Employing \eqref{res_029} and defining 
$\mathcal K_{\rm cs} = -\log \left[ \: i \int\Omega\wedge \ov\Omega  \:\right]$, we arrive at
\eq{
  \label{rel_02}
    \int A\wedge \star \ov B =  e^{\mathcal K_{\rm cs}} \biggl[ \hspace{20pt} &
    G^{\mu\ov\nu} D_{U^{\mu}} \bigl( \textstyle{\int} A\wedge \Omega)\;
    D_{\ov U^{\ov\nu}} \bigl( \textstyle{\int} \ov B\wedge \ov\Omega) \\
    +\,&G^{\mu\ov\nu} D_{U^{\mu}} \bigl( \textstyle{\int} \ov B\wedge \Omega)\;
    D_{\ov U^{\ov\nu}} \bigl( \textstyle{\int} A\wedge \ov\Omega) \\[6pt]
    +\,& \bigl( \textstyle{\int} A\wedge \ov\Omega)\; \bigl( \textstyle{\int} \ov B \wedge \Omega) \\[0pt]
    +\,& \bigl( \textstyle{\int} A\wedge \Omega)\; \bigl( \textstyle{\int} \ov B \wedge \ov\Omega)   
    \hspace{80pt}\biggr] \,.
}
Similarly, we determine for the wedge product of two three-forms $A$ and $B$
\eq{
  \label{rel_022}
    \int A\wedge \ov B = -i\, e^{\mathcal K_{\rm cs}} \biggl[ \hspace{20pt} &
    G^{\mu\ov\nu} D_{U^{\mu}} \bigl( \textstyle{\int} A\wedge \Omega)\;
    D_{\ov U^{\ov\nu}} \bigl( \textstyle{\int} \ov B\wedge \ov\Omega) \\
    -\,&G^{\mu\ov\nu} D_{U^{\mu}} \bigl( \textstyle{\int} \ov B\wedge \Omega)\;
    D_{\ov U^{\ov\nu}} \bigl( \textstyle{\int} A\wedge \ov\Omega) \\[6pt]
    +\,& \bigl( \textstyle{\int} A\wedge \ov\Omega)\; \bigl( \textstyle{\int} \ov B \wedge \Omega) \\[0pt]
    -\,& \bigl( \textstyle{\int} A\wedge \Omega)\; \bigl( \textstyle{\int} \ov B \wedge \ov\Omega)   
   \hspace{80pt}\biggr] \,.
}

%%%%%%%%%%%%%%%%%%%%%%%%%%%%%%%%%%%%%%%%%%%%%%%
%%%%%%%%%%%%%%%%%%%%%%%%%%%%%%%%%%%%%%%%%%%%%%%

\subsection{Relations regarding $\mathcal D$}

We now derive relations for the twisted differential $\mathcal D$, which was
defined via \eqref{deffluxes}. Let us consider a closed three-form $\mathsf A$ with $d\mathsf A=0$,
and expand $\mathsf A$ the basis \eqref{basis_001} as
\eq{
  \mathsf A = \mathsf A^{\Lambda} \alpha_{\Lambda} + \mathsf A_{\Lambda} \beta^{\Lambda} \,.
}
Using the definitions \eqref{deffluxes}, we can then show by explicit
computation that
\eq{
  \label{res_035}
  \arraycolsep2pt
  \begin{array}{lccl}
  \displaystyle \int \mathcal D\op \omega_A\wedge \mathsf A  &=& 
  -&\displaystyle \int \omega_A\wedge\mathcal D \mathsf A \,, \\[10pt]
  \displaystyle \int \mathcal D\op \sigma^A  \wedge \mathsf A &=& 
  +& \displaystyle  \int \sigma^A\wedge \mathcal D \mathsf A \,.
  \end{array}
}
Let us also consider an even, $d$-closed  multi-form $\mathsf B$, which can be  
expanded in the basis \eqref{res_033} as
\eq{
 \mathsf B=\mathsf B^A\omega_A + \mathsf B_A \sigma^A\,.
} 
For a Calabi-Yau three-fold with the action of $\mathcal D$ given by \eqref{deffluxes}, it follows that
$\mathcal D\op\mathsf B$ is a three-form. 
Setting then $\mathsf A = \mathcal D\op\mathsf B$ and using the Bianchi identities 
$\mathcal D^2=0$, it follows that 
\eq{
  \label{res_034}
  \int \mathcal D\op  \mathsf B \wedge \mathcal D\op \omega_A  = 0 \,, \hspace{50pt}
  \int \mathcal D\op  \mathsf B \wedge \mathcal D\op \sigma^A  = 0  \,.
}

%%%%%%%%%%%%%%%%%%%%%%%%%%%%%%%%%%%%%%%%%%%%%%%
%%%%%%%%%%%%%%%%%%%%%%%%%%%%%%%%%%%%%%%%%%%%%%%

\subsection{K\"ahler metric and inverse}

We now discuss the K\"ahler metric $G_{i\ov j}$ for the moduli $\tau$, $G^a$ and $T_{\alpha}$, which were 
defined in \eqref{moduli}. 
From \cite{Benmachiche:2006df} we know that this metric can be expressed as
\eq{
  \label{met}
  G_{i\ov j} = \frac{e^{2\phi}}{4\op \mathcal V} \int \bigl[ \nu_i \wedge e^{+B} \bigr]
  \wedge \star  \bigl[ \nu_j \wedge e^{+B} \bigr] \,,
}
where $i,j=\tau,\op  G^a,\op T_{\alpha}$ and 
\eq{
  \label{con_73}
  \nu_i = \bigl( \: 1 \,,\, -\omega_a \,,\, \sigma^{\alpha}\bigr) \,,
}
and where  $\mathcal V$ denotes the volume of the Calabi-Yau three-fold in string frame.
The inverse K\"ahler metric has not been given in \cite{Benmachiche:2006df}, but can be determined 
as follows. Let us make the following ansatz
\eq{
  \label{inv_met}
  G^{i\ov j} = \frac{4\op \mathcal V}{e^{2\phi}} \int \bigl[ \rho^i \wedge e^{-B} \bigr]
  \wedge \star  \bigl[ \rho^j \wedge e^{-B} \bigr] \,,
}
with the dual forms
\eq{
  \rho^i = \bigl( \: \omega_0 \,,\,-\sigma^a \,,\,  \omega_{\alpha}\bigr) \,.
}
We now verify that \eqref{inv_met} is indeed the inverse of \eqref{met}. 
For that purpose, we note that
\eq{
  \int \bigl[ \nu_i \wedge e^{+B} \bigr] \wedge \bigl[ \rho^j \wedge e^{-B} \bigr] =
  \int \nu_i \wedge \rho^j = 
  \delta_i^j \,.
}
This implies that we can expand the Hodge duals as
\eq{
  \star\bigl( \nu_i \wedge e^{+B} \bigr) = \mathcal M_{ij}   \, \bigl( \rho^j \wedge e^{-B} \bigr) \,,
  \hspace{40pt}
  \star\bigl( \rho^i \wedge e^{-B} \bigr) = \mathcal N^{ij}   \, \bigl( \nu_j \wedge e^{+B} \bigr) \,,
}
with $\mathcal M$ and $\mathcal N$ some matrices. 
Applying the Hodge star to the second relation and noting that for even forms on six-dimensional
manifold $\star^2=1$, gives
$\mathcal N^{ij} \op\mathcal M_{jk} = \delta^i_k$. This allows us to
compute
\eq{
  G_{i\ov j} \,G^{\ov j k} = \mathcal M_{ji}\, \mathcal N^{kj}
  = \mathcal N^{kj}\, \mathcal M_{ji}  = \delta^k_i \,.
}
We have therefore shown that  the metric \eqref{inv_met} is indeed the inverse of \eqref{met}.

%%%%%%%%%%%%%%%%%%%%%%%%%%%%%%%%%%%%%%%%%%%%%%%
%%%%%%%%%%%%%%%%%%%%%%%%%%%%%%%%%%%%%%%%%%%%%%%
%%%%%%%%%%%%%%%%%%%%%%%%%%%%%%%%%%%%%%%%%%%%%%%
%%%%%%%%%%%%%%%%%%%%%%%%%%%%%%%%%%%%%%%%%%%%%%%

\section{Proof of general results}
\label{app_proof}

In this appendix we show that for vanishing $B$-field, $\star\op \mathcal L_{\rm NS\op NS}$ can indeed be 
cast as proposed in \eqref{final3}.
We have already proved that \eqref{res_006}, \eqref{res_007}, \eqref{res_009}, \eqref{res_010} and \eqref{resOR},
comply with \eqref{final3} when only one kind of flux is switched on at a time. When fluxes are turned on simultaneously
we have to care about mixed terms. In the original NS-NS Lagrangian there are $FR$ and $HQ$ mixed terms in \eqref{lag_2}. 
On the other hand, given its structure, in \eqref{final3} the only mixed terms are precisely of type 
$FR$ and $HQ$. Concretely, the relevant terms in \eqref{final3} are $T_{HQ} + T_{FR}$, where
\eq{
\label{resHQ}
T_{HQ} =
-H\wedge\star (Q\bullet \tfrac{1}{2}J^2)
 + {\rm Re} (\Omega\wedge H) \wedge \star (\ov\Omega\wedge Q\bullet \tfrac{1}{2}J^2) \, ,
}
and similarly
\eq{
\label{resFR}
T_{FR}=
-F\circ J \wedge\star(R \,\llcorner  \tfrac{1}{3!} \op J^3 )
 + {\rm Re} (\Omega\wedge F\circ J) \wedge \star (\ov\Omega\wedge R \,\llcorner  \tfrac{1}{3!}J^3) \, .
}
We will proceed by evaluating separately each term in the above relations.

Let us begin with \eqref{resHQ}. Using \eqref{resQJ2} and the property $J_{ij}=g_{im} \, I^m{}_j$ we find
\eq{
\label{resHQ11}
-H\wedge\star (Q\bullet \tfrac{1}{2}J^2) = -\frac 12 H_{i'j'k'} Q_i^{\ jk}
I^{j'}{}_j\op I^{k'}{}_{k}\op  g^{ii'} \op \star 1 \, .
}
It is convenient to express the right hand side in a complex basis and then simplify it
applying an appropriate Bianchi identity. With $F$, $H$ and $Q$ different from zero, the second identity
in \eqref{bianchids1} yields
\begin{eqnarray}
  \label{BIFHQ}
 g^{a\ov a}\big(H_{abc} Q_{\ov a}^{\ bc}+  H_{\ov a \ov b \ov c} Q_{a}^{\ \ov b \ov c}\big)  & - &
 g^{a\ov a}\big(H_{\ov a bc} Q_{a}^{\ bc}+  H_{a \ov b \ov c} Q_{\ov a}^{\ \ov b \ov c} \big) 
 \nonumber \\[2mm]
&+ & 2  g^{a\ov a} \big( F^{\ov c}{}_{a b}\,  F^{b}{}_{\ov a \ov c} -
F^c{}_{a\ov b}\,  F^{\ov b}{}_{\ov a c} \big) = 0 
  \, .
\end{eqnarray}
Notice that when only $F\not= 0$ this identity reduces to \eqref{res_005}.
Going to a complex basis and substituting \eqref{BIFHQ} we arrive at
\begin{eqnarray}
\label{resHQ12}
-H\wedge\star (Q\bullet \tfrac{1}{2}J^2) = \hspace*{-5mm} & & \hspace*{-3mm} 
  g^{a\ov a}\big(H_{abc} Q_{\ov a}^{\ bc}+  H_{\ov a \ov b \ov c} Q_{a}^{\ \ov b \ov c} -
 H_{\ov ab \ov c} Q_{a}^{\ b \ov c} - H_{a  b \ov c} Q_{\ov a}^{\  b \ov c} \big) \op \star 1 
\nonumber \\[2mm]
& {-} & \hspace*{-1mm} 
g^{a\ov a} \big( F^{\ov c}{}_{a b}\,  F^{b}{}_{\ov a \ov c} -
F^c{}_{a\ov b}\,  F^{\ov b}{}_{\ov a c} \big) \op \star 1 
 \, .
\end{eqnarray}
The $F$ depending piece will cancel against an analogous contribution in $\frac{1}{2} \, \Xi_3\wedge \star\op \Xi_3$, 
$\Xi_3 = F\circ J$. In fact, {} from \eqref{res_003} we see that before using \eqref{res_005}, the right hand side of 
\eqref{xithree} has an extra term that offsets the second line in \eqref{resHQ12}.
In the complex basis we also obtain
\eq{
\label{resHQ2}
{\rm Re} (\Omega\wedge H) \wedge \star (\ov\Omega\wedge Q\bullet \tfrac{1}{2}J^2) =
-2  g^{a\ov a}\big(H_{abc} Q_{\ov a}^{\ bc}+  H_{\ov a \ov b \ov c} Q_{a}^{\ \ov b \ov c}\big) \op \star 1
\, .
}
Finally, for the $HQ$ term in \eqref{lag_2} the Bianchi identity \eqref{BIFHQ} further implies that
\begin{eqnarray}
\label{resHQ0}
-\frac12 H_{mni} Q_j^{\ mn} g^{ij} \op\star 1=  & -& \hspace*{-1mm}
 g^{a\ov a}\big(H_{abc} Q_{\ov a}^{\ bc}+  H_{\ov a \ov b \ov c} Q_{a}^{\ \ov b \ov c} +
 H_{\ov ab \ov c} Q_{a}^{\ b \ov c} + H_{a  b \ov c} Q_{\ov a}^{\  b \ov c} \big) \op\star 1
\nonumber \\[2mm]
& {+} & \hspace*{-1mm} 
g^{a\ov a} \big( F^{\ov c}{}_{a b}\,  F^{b}{}_{\ov a \ov c} -
F^c{}_{a\ov b}\,  F^{\ov b}{}_{\ov a c} \big) \op\star 1
 \, .
\end{eqnarray}
The term involving $F$ is cancelled by a similar contribution in  $\frac12 F^m_{\ ni}F^n_{\ mj} g^{ij} \op\star 1$
that also appears in \eqref{lag_2}. In the analysis of pure $F$ flux this extra contribution was absent by virtue of \eqref{res_005}. 
Observe that adding the first line in \eqref{resHQ12} and \eqref{resHQ2} precisely matches the first line in \eqref{resHQ0}.
Hence, we have shown that the mixed terms in $T_{HQ}$ indeed lead to the $HQ$ term in the NS-NS Lagrangian.

To evaluate the mixed $FR$ terms we basically take the same steps as in the preceding calculation.
A crucial ingredient is the Bianchi identity that follows from the fourth line in \eqref{bianchids1}
\begin{eqnarray}
  \label{BIQRF}
 g_{a\ov a}\big(R^{\ov a bc} F^{a}_{\ bc}+  R^{a \ov b \ov c} F^{\ov a}_{\ \ov b \ov c} \big)  & - &
 g_{a\ov a}\big(R^{abc} F^{\ov a}_{\ bc}+  R^{\ov a \ov b \ov c} F^{a}_{\ \ov b \ov c}\big)  
 \nonumber \\[2mm]
&+ & 2  g_{a\ov a} \big( Q_{\ov b}{}^{a c}\,  Q_{c}{}^{\ov a \ov b} -
Q_b^{\ a\ov c}\,  Q_{\ov c}^{\ \ov a b} \big) = 0   \, ,
\end{eqnarray}
which clearly shortens to \eqref{BIQQ} when only $Q\not=0$. Inserting this identity in the $FR$ term in \eqref{lag_2} gives
\begin{eqnarray}
\label{resFR0}
-\frac12 R^{mni} F^j_{\ mn} \op g_{ij} \star 1=  & -& \hspace*{-1mm}
 g_{a\ov a}\big(R^{abc} F^{\ov a}_{\ bc}+  R^{\ov a \ov b \ov c} F^{a}_{\ \ov b \ov c} +
 R^{\ov a b \ov c} F^{a}_{\ b \ov c} + R^{a  \ov b c} F^{\ov a}_{\  \ov b c} \big) \op\star 1
\nonumber \\[2mm]
& {+} & \hspace*{-1mm} 
g_{a\ov a} \big( Q_{\ov c}^{a b}\,  Q_{b}{}^{\ov a \ov c} -
Q_c^{a\ov b}\,  Q_{\ov b}{}^{\ov a c} \big) \op\star 1
 \, .
\end{eqnarray}
The $Q$ part is nullified by an identical term with opposite sign in  \mbox{$\frac12 Q_m^{\ ni} Q_n^{\ mj} g_{ij} \op\star 1$}.
Using the identity \eqref{BIQRF} we also find that adding the pieces in $T_{FR}$ reproduces the first line in \eqref{resFR0}
up to an additional contribution that is cancelled by a similar one in $Q\bullet\frac12 J^2 \wedge \star (Q\bullet\frac12 J^2)$.

%%%%%%%%%%%%%%%%%%%%%%%%%%%%%%%%%%%%%%%%%%%%%%%
%%%%%%%%%%%%%%%%%%%%%%%%%%%%%%%%%%%%%%%%%%%%%%%
%%%%%%%%%%%%%%%%%%%%%%%%%%%%%%%%%%%%%%%%%%%%%%%
%%%%%%%%%%%%%%%%%%%%%%%%%%%%%%%%%%%%%%%%%%%%%%%
%%%%%%%%%%%%%%%%%%%%%%%%%%%%%%%%%%%%%%%%%%%%%%%
%%%%%%%%%%%%%%%%%%%%%%%%%%%%%%%%%%%%%%%%%%%%%%%
%%%%%%%%%%%%%%%%%%%%%%%%%%%%%%%%%%%%%%%%%%%%%%%
%%%%%%%%%%%%%%%%%%%%%%%%%%%%%%%%%%%%%%%%%%%%%%%
 
\clearpage
\bibliography{references}  
\bibliographystyle{utphys}

%%%%%%%%%%%%%%%%%%%%%%%%%%%%%%%%%%%%%%%%%%%%%%%
%%%%%%%%%%%%%%%%%%%%%%%%%%%%%%%%%%%%%%%%%%%%%%%
%%%%%%%%%%%%%%%%%%%%%%%%%%%%%%%%%%%%%%%%%%%%%%%
%%%%%%%%%%%%%%%%%%%%%%%%%%%%%%%%%%%%%%%%%%%%%%%

\end{document}